\documentclass[%
 reprint,
 amsmath,amssymb,
 aps,
pra,
]{revtex4-1}

\DeclareMathAlphabet\EuScript{U}{eus}{m}{n}
\SetMathAlphabet\EuScript{bold}{U}{eus}{b}{n}
\pdfoutput=1

\usepackage{graphicx}
\usepackage{dcolumn}
\usepackage{bm}

\begin{document}


\title{Measuring Nuclear Spin Dependent Parity Violation With Molecules: Experimental Methods and Analysis of Systematic Errors} 


\author{Emine Altunta\c{s}}
\email[]{emine.altuntas@yale.edu}
\altaffiliation[Present Address: ] {Joint Quantum Institute, National Institute of Standards and Technology, and University of Maryland, Gaithersburg, MD 20899}
\author{Jeffrey Ammon}
\altaffiliation[Present Address: ] {Lincoln Laboratory, Massachusetts Institute of Technology, Lexington, MA 02420}
\author{Sidney B. Cahn}
\email[]{sidney.cahn@yale.edu}
\affiliation{Department of Physics, P.O.Box 208120, Yale University, New Haven, Connecticut 06520, USA}
\author{David DeMille}
\email[]{david.demille@yale.edu}
\affiliation{Department of Physics, P.O.Box 208120, Yale University, New Haven, Connecticut 06520, USA}


\date{\today}

\begin{abstract}
Nuclear spin-dependent parity violation (NSD-PV) effects in atoms and molecules arise from $Z^0$ boson exchange between electrons and the nucleus, and from the magnetic interaction between electrons and the parity-violating nuclear anapole moment. It has been proposed to study NSD-PV effects using an enhancement of the observable effect in diatomic molecules [D. DeMille \textit{et al.}, Phys. Rev. Lett. \textbf{100}, 023003 (2008)]. Here, we demonstrate measurements of this type with sensitivity surpassing that of any previous atomic PV measurement, using the test system ${^{138}\mathrm{Ba^{19}F}}$. We show that systematic errors associated with our technique can be suppressed to at least the level of the present statistical sensitivity. With $\sim\!170$ hours of data, we measure the matrix element, $W$, of the NSD-PV interaction with uncertainty $\delta W<0.7$ Hz, for each of two configurations where $W$ must have different signs. This sensitivity would be sufficient to measure NSD-PV effects of the size anticipated across a wide range of nuclei. 
\end{abstract}


\maketitle

\section{INTRODUCTION}

The weak force leads to interactions that are not the same under inversion of spatial coordinates, i.e., parity (P) is violated. In atoms and molecules, certain aspects of the parity violating (PV) electroweak interactions are manifested by nuclear spin dependent parity violation (NSD-PV). NSD-PV leads to small and poorly-characterized effects that primarily arise from two fundamental causes. The first is the electroweak coupling between vector-electron and axial-nucleon neutral currents ($\ensuremath{V_eA_n}$) that results from $Z^0$ boson exchange between electrons and nucleons. So far, $\ensuremath{V_eA_n}$ measurements at low momentum transfer have come from electron-nucleus scattering experiments; their results are typically expressed in terms of the dimensionless constants, $C_{2u,d}$, that characterize the $\ensuremath{V_eA_n}$ coupling to the up and down quarks~\cite{Slac, Pvdis, Sample, ERLER2013119}. However, $C_{2u,d}$ are numerically small in the Standard Model, and their present experimental uncertainties are relatively large, $\gtrsim\!70\%$ of the predicted values~\cite{Sample}. Improvements in $C_{2u,d}$ measurements would provide a new check of the Standard Model. The second source of NSD-PV is the nuclear anapole moment, which arises from weak interactions within the nucleus~\cite{Zeldovich}. This P-odd magnetic moment can couple to the magnetic dipole carried by the spin of a penetrating electron and results in a contact potential between the electron and the nucleus~\cite{Flambaum}. A nonzero nuclear anapole moment has been measured only once, in $^{133}\mathrm{Cs}$~\cite{Wieman}. Additional measurements of anapole moments may enable determination of parameters describing the strength of purely hadronic PV interactions~\cite{Murray, Haxton}, which have proven difficult to measure by other means~\cite{Gardner,HAXTON2013185}. It has been suggested that anapole moment measurements might provide a benchmark for assessing the accuracy of calculated values for nuclear matrix elements describing neutrinoless double beta decay~\cite{Ramsey-Musolf}.
 
There are several current experiments using atomic physics techniques to investigate NSD-PV in isotopes of Dy, Yb, Ra, Fr, and Cs, which are complementary to the work reported here. The most sensitive atomic PV experiment (prior to the results presented here) was performed in atomic Dy~\cite{Nguyen}, and yielded a null measurement of the PV effect, though this particular measurement was sensitive only to nuclear spin independent (NSI) PV effects. Subsequently, the theoretical expectation for the size of PV effects in Dy was reduced~\cite{Dzuba}, and a new experiment~\cite{Leefer} is planned to measure NSD-PV in Dy. Another experiment~\cite{Antypas2017} is planned to measure NSD-PV in Yb, in which a predicted enhancement of PV effects \cite{DeMille95} has been observed~\cite{TsigutkinYb09,TsigutkinYb10}. As was the case with the older experiments using $^{133}\mathrm{Cs}$~\cite{Wieman} and Tl~\cite{VetterFortson} (where a significant upper bound on NSD-PV effects was set), these experiments, as well as a new one in Ra$^+$~\cite{NunezPortela2013} require measurements using several different hyperfine-resolved transitions to separate NSD-PV effects from the much larger NSI-PV effects. By contrast, new experiments in Cs~\cite{Antypas2013}, and Fr~\cite{Aubin2013} are aimed at a direct measurement of NSD-PV effects.

In this paper, we demonstrate a statistical sensitivity to NSD-PV surpassing that of any previous atomic PV measurement. Our experiment is based on the novel approach of using diatomic molecules. Diatomic molecules with a single valence electron in a $X^2\Sigma$ ground state offer several important traits that we employ in our method to measure NSD-PV effects~\cite{Labzowsky95, FLAMBAUM1985121,FlambaumSushkov}. Due to their rotational/hyperfine structure, diatomic molecules systematically have about 5 orders of magnitude smaller energy splittings between opposite parity states that can be mixed by NSD-PV than are typically found in atoms. Magnetic fields of a technically feasible magnitude can be used to Zeeman-shift these molecular levels to near degeneracy. This near-degeneracy enhances the mixing of opposite-parity states due to NSD-PV, and leads to much larger PV signals than can be obtained in more traditional approaches to measuring PV in atoms. To measure the strength of the mixing induced by NSD-PV effects, a Stark interference technique to enhance the observable effect due to the NSD-PV induced mixing  has been proposed~\cite{DeMille}, similar to that used in experiments with atomic Dy~\cite{Nguyen}.
 
We report here on measurements of NSD-PV using the molecule BaF. BaF is a prototypical case of the type of molecule where the general approach described above can be employed. Its properties (described below) make it relatively easy to work with, from a technical point of view. BaF has both odd and even nucleon number stable isotopes with good natural abundances. NSD-PV effects are nonzero only for isotopes with a nonzero nuclear spin, $I$, i.e., with an unpaired proton or neutron \cite{DeMille}. For the isotopomer $^{137}\mathrm{BaF}$, the predicted size of the NSD-PV effect associated with $^{137}$Ba is well above the sensitivity we demonstrate here. 

Here, however, we use $^{138}\mathrm{BaF}$ to demonstrate our measurement technique. This was done for a few reasons. First, the raw signals from $^{138}\mathrm{BaF}$ are $\sim\!20\times$ larger than for $^{137}\mathrm{BaF}$; this is because $^{138}\mathrm{BaF}$ has a higher abundance than $^{137}\mathrm{BaF}$ ($72\%$ vs. $11\%$) and fewer nuclear-spin projection sublevels ($I_{Ba}=0$ for $^{138}\mathrm{BaF}$ and $I_{Ba}=3/2$ for $^{137}\mathrm{BaF}$), only one of which can be used during a measurement. Since $I_{\rm Ba}=0$ for $^{138}\mathrm{BaF}$, here NSD-PV effects are due only to $^{19}{F}$, with $I_{F}=1/2$. However, since the valence electron wavefunction in BaF has only very small overlap with the F nucleus, the anticipated effect due to $I_F$ is far below our experimental sensitivity. As such, $^{138}\mathrm{BaF}$ is a powerful system with which to identify and measure contributions due to systematic errors: any apparent NSD-PV signal in the measurements we report here could be due to \textit{only} systematic errors. Hence, the primary result of this paper is to demonstrate control over systematic errors at a level sufficient for future measurements in molecular species (including $^{137}$BaF) where NSD-PV effects are nonzero.

\subsection{Relevant Molecular Structure of $^{138}$BaF}
\label{138BaF}
The ground electronic state $X^2\Sigma$ of $^{138}\mathrm{BaF}$ is described by the effective Hamiltonian $H = {B_e\bm{N^2}} + {\gamma\bm{N\cdot S}} + {b \bm{I\cdot S}} + c(\bm{I\cdot n})(\bm{S\cdot n})$, where $N$ is the rotational angular momentum, $S=1/2$ is the electron spin, $B_e$ is the rotational constant, $\gamma$ is the spin-rotation (SR) constant, $b,c$ are hyperfine (HF) constants and $\bm{n}$ is a unit vector along the internuclear axis ($\hbar=1$ throughout). Here $N$ is a good quantum number, with eigenstates of energy $E_N \approx B_eN(N+1)$ and parity $P=(-1)^N$. We Zeeman shift sublevels of the $N^P=0^+$ and $1^-$ manifolds of states to near degeneracy, using a magnetic field $\boldsymbol{\EuScript{B}}=\EuScript{B}\hat{z}$. Zeeman shifts are dominated by the coupling to $\bm{S}$, with approximate Hamiltonian $H_Z\cong-g\mu_B\bm{S}\cdot\boldsymbol{\EuScript{B}}$, where $g\cong-2$ and $\mu_B$ is the Bohr magneton. Since $B_e\gg\gamma,b,c$, the $\mathcal{B}$-field necessary to bridge the rotational energy ${E_1}-{E_0}\approx\!2B_e$ is large enough to strongly decouple $\bm{S}$ from $\bm{I}$ and $\bm{N}$. Thus we write the molecular states in the decoupled basis $| N, m_N \rangle |S, m_s \rangle | I, m_I \rangle$, which are good approximations to the energy eigenstates near the level crossings. Level crossings between pairs of opposite-parity states with different values of $(m_N, m_I)$ and $(m_N^{'}, m_I^{'})$ occur at different values of $\mathcal{B}$ because of energy differences in the sublevels due to hyperfine and spin-rotation terms in $H$~\cite{DeMille}.  

\begin{figure}
\includegraphics[width=83mm]{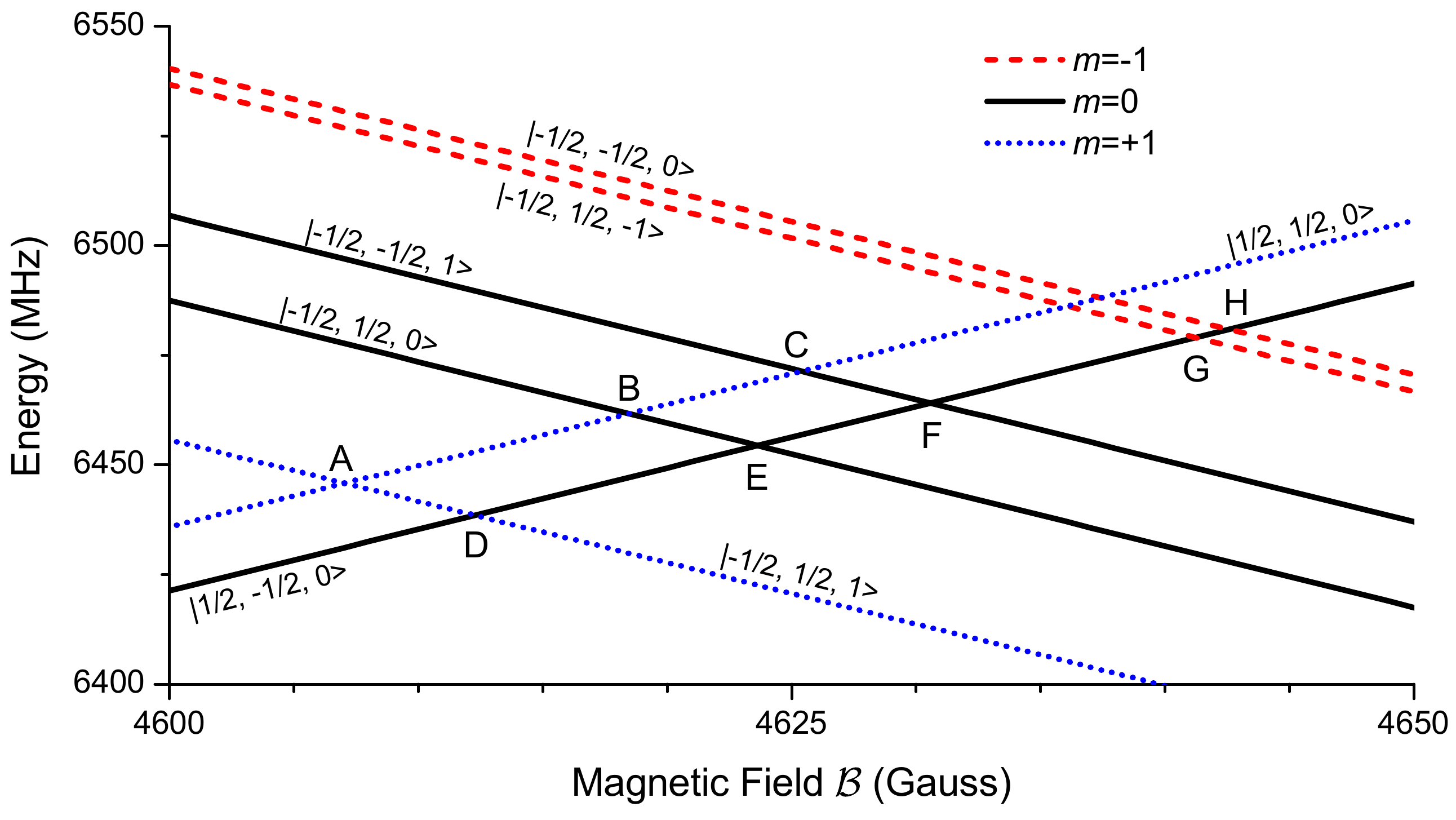}
\caption{(color online). Level crossings in $^{138}\mathrm{BaF}$. Up- (down-) sloping levels belong to the even- (odd-) parity ${N=0~(N=1)}$ rotational level. Kets label the approximate quantum numbers $|m_S,m_I, m_N\rangle$, and the legend shows the value of $m = m_S + m_I + m_N$. Letters label each crossing where levels can be mixed via the Stark effect.  NSD-PV leads to mixing only between  levels with the same value of $m$. We measure NSD-PV at crossings A and F. Reproduced from~\cite{CahnPRL}.}
\label{fig:Crossings138}
\end{figure}

In $^{138}\mathrm{BaF}$, the levels $ |0,0\rangle|\frac{1}{2},\frac{1}{2}\rangle|\frac{1}{2},m_I\rangle\equiv| \psi_{\uparrow}^{+} (m_N=0, m_I)\rangle$ (even parity) and $|1,m_N^{'}\rangle|\frac{1}{2},-\frac{1}{2}\rangle|\frac{1}{2},m_I^{'}\rangle\equiv
|\psi_{\downarrow}^{-} (m_N^{'}, m_I^{'})\rangle$ (odd) are degenerate under $H+H_Z$ when $\mathcal{B}={\mathcal{B}}_0 \approx\! B_e/\mu_B\sim0.5$ T. Fig.~\ref{fig:Crossings138} shows the level crossings in $^{138}\mathrm{BaF}$. The NSD-PV Hamiltonian is a pseudoscalar that mixes levels with opposite parity and the same value of the total angular momentum projection $m$, where $m = m_S + m_N + m_I$. We use the crossings labeled A (where $m_I =  m_I^\prime = 1/2$, $m_N^\prime = 1$, and $m$ = 1) and F (where $m_I = m_I^\prime = -1/2$, $m_N^\prime = 1$, and $m = 0$) in our NSD-PV measurements here~\cite{CahnPRL}.

For a diatomic molecule in a $^2\Sigma$ state, the effective Hamiltonian $H^{\rm eff}_{\rm P}$ of the NSD-PV interaction is~\cite{Flambaum1985} 
\begin{equation}
\label{hamiltonian2}
H^{\rm eff}_{\rm P} =\kappa' W_P \frac{\left(\bm{n} \times \bm{S}\right)\cdot \bm{I}}{I}.
\end{equation}
Here, $\kappa^\prime$ is a dimensionless
number parameterizing the strength of the
NSD-PV interaction; it depends only on the physics of electroweak interactions and nuclear structure, and is independent of any molecular parameters. The parameter $W_P$ characterizes the strength of the matrix element (ME) of the parity violating Hamiltonian in the ${^2}\Sigma$ wavefunction of the electron, for a given nucleus, in a frame where the molecule is not rotating; it is independent of electroweak and nuclear effects, and can be calculated accurately for many different molecular species. All the angular momentum dependence of $H^{\rm eff}_{\rm P}$ is encoded in the dimensionless operator ${C\equiv\left(\bm{n} \times \bm{S}\right)\cdot \bm{I}/{I}}$. Then the NSD-PV ME of mixed nearly-degenerate levels is 
\begin{equation}
iW(m_N^{'},m_I^{'},m_N,m_I)\equiv\kappa' W_P \tilde{C},
\label{Eqn.W2}
\end{equation}
where $iW$ is pure imaginary due to time-reversal invariance. Here, the ME $\tilde{C} \equiv \langle \psi_{\downarrow}^{-}(m_N^{'}, m_I^{'}) |C |\psi_{\uparrow}^{+} (m_N, m_I) \rangle$ depends only on the angular momentum content of the near-degenerate states, while the quantity $W_{\rm mol}\equiv\kappa^\prime W_P$ is the same at all crossings in a given molecule and for a given nucleus. $W_P$ has been calculated several times for Ba in BaF, and also recently for F in BaF, with values $W_P(\rm {Ba}) = 164$ Hz and $W_P(\rm{F}) = 0.05$ Hz~\cite{DeMille, Private}. The parameter $\kappa^\prime = \kappa_2^\prime + \kappa_a^\prime$ has contributions from the $V_eA_n$ interaction ($\kappa_2^\prime$) and from the electron-nuclear anapole moment interaction ($\kappa_a^\prime$). The value of $\kappa^\prime$ encodes the physics of the NSD-PV interaction, and is ultimately the quantity we seek to measure.

\subsection{Measurement Principle: Idealized Case}
\label{IdealMeasure}

\begin{figure}
\includegraphics[width=83mm]{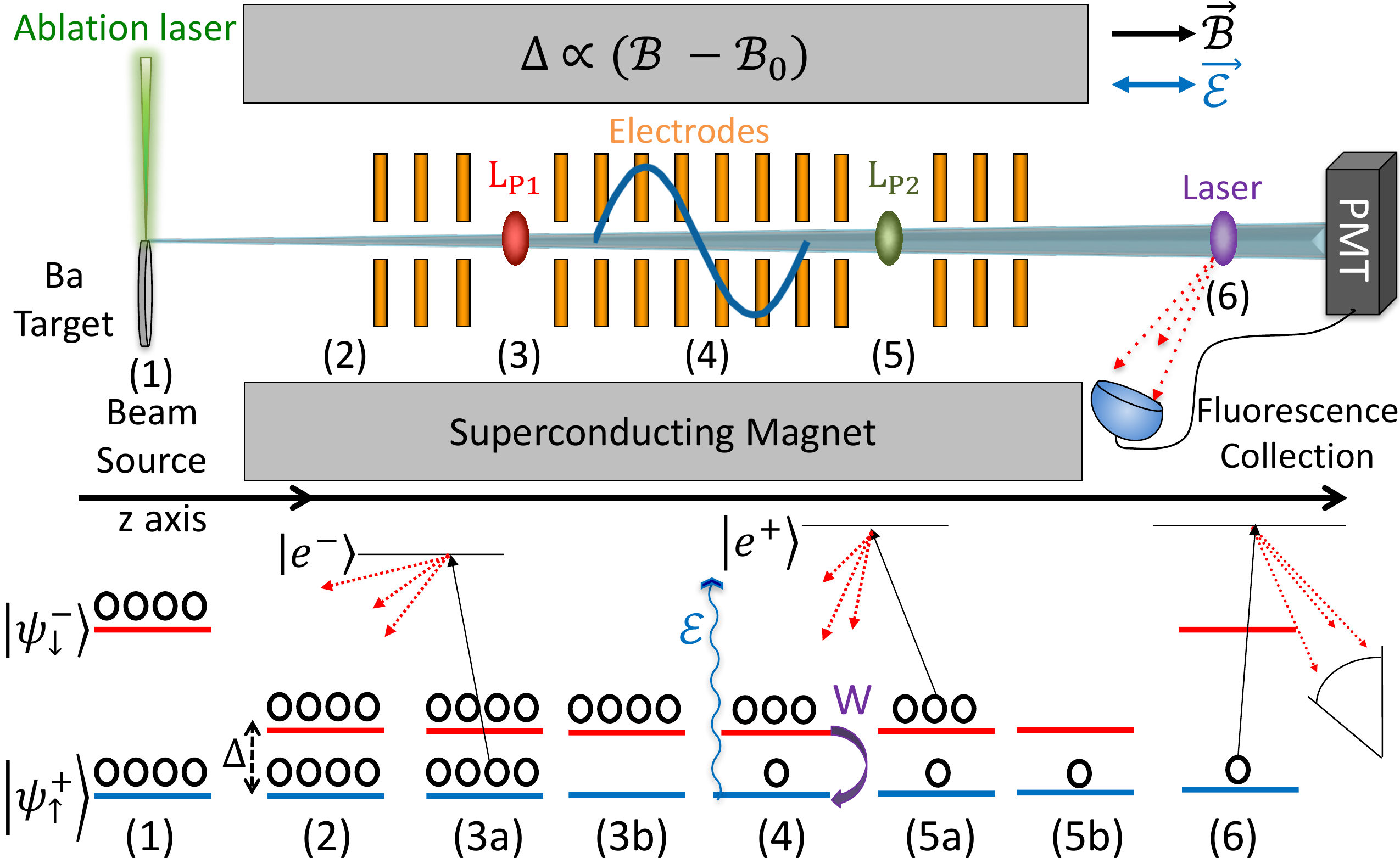}
\caption{(color online).  Schematic of the apparatus (top) and evolution of the level populations (bottom). (1) BaF molecules are formed by laser ablation into a pulsed jet; both parity states have equal thermal populations. (2) Molecules enter the magnet and levels are brought to near-degeneracy. (3) Laser beam $\rm{L}_{P1}$ (red) depletes the even-parity $|\psi^+_\uparrow\rangle$ state by optical pumping via an odd-parity excited state, $|e^-\rangle$. (4) A single-cycle sine wave $\mathcal{E}$-field (blue) is applied parallel to the $\boldsymbol{\EuScript{B}}$-field. Stark and NSD-PV interactions combine to mix opposite parity states and transfer population into $|\psi^+_\uparrow\rangle$. (5) Laser beam $\rm{L}_{P2}$ (green) depletes the odd-parity $|\psi^-_\downarrow\rangle$ state by optical pumping via an even-parity state, $|e^+\rangle$. (6) Molecules exit the magnet, and population transferred to $|\psi^+_\uparrow\rangle$ by the $\mathcal{E}$-field and the NSD-PV interaction is detected by laser-induced (purple) fluorescence. }
\label{fig:ApparatusSchematic}
\end{figure}

We use a Stark-interference method to measure $iW$. Fig.~\ref{fig:ApparatusSchematic} shows a schematic of the experiment and evolution of molecular levels during the measurement. We first discuss an idealized version of the measurement protocol. A beam of $^{138}\mathrm{BaF}$ molecules enters the magnet with field $\EuScript{B}\approx\! \EuScript{B}_0$. Next, we deplete the even-parity state $|\psi^+_\uparrow\rangle$, by optically pumping to unobserved ground-state sublevels via a short-lived electronic state of definite negative parity, $|e^-\rangle$. This state preparation occurs at a time defined as $t=0$. In this idealized protocol, the molecules then immediately enter a region with a spatially varying electric field, $\boldsymbol{\mathcal{E}} = \mathcal{E}_0 \sin(2\pi z/L)\hat{z}$ for $0<z<L$. The Hamiltonian for the near-degenerate states, written in the basis of parity eigenstates, is
\begin{align}
H_{\pm} =& \left( \begin{array}{c} 0 \\ 
-iW+d\mathcal{E}(t) \end{array} \right. & \left.
\begin{array}{c} iW+d\mathcal{E}(t) \\ \Delta\end{array}\right),
\label{Hplusminus}
\end{align}
where $\Delta$ is the small $\mathcal{B}$-field dependent detuning from exact degeneracy, and $d$ is the dipole matrix element~\cite{DeMille}. Though the electric field is static in time, its spatial dependence causes molecules with velocity $\textbf{v}=v\hat{z}$ to experience a time-dependent field $\mathcal{E}(t=z/v) = \mathcal{E}_0 \sin(\omega t)$, with $\omega=2\pi v/L$. The wavefunction is 
\begin{equation}
| \psi\left(t\right)\rangle = c_+\left(t\right)|\psi_{\uparrow}^{+}\rangle + e^{-i\Delta t} c_-\left(t\right)|\psi_{\downarrow}^{-}\rangle \equiv \left(\begin{array}{c} c_+\left(t\right) \\ c_-\left(t\right)\end{array}\right),
\label{Equ:Shro}
\end{equation}
where $c_+\left(0\right) =0$ and $c_-\left(0\right) =1$ due to the optical pumping. Then, solving the Schr\"{o}dinger equation with the assumption $W\ll d\mathcal{E}_0$, we find
\begin{equation}
\begin{split}
c_+\left(t\right) \approx \!& -2 i e^{-i\Delta t/2}  \biggl[\cos\left(\frac{\Delta t}{2}\right)\frac{d\mathcal{E}_0\omega}{\omega^2-\Delta^2}\sin^2\left(\frac{\omega t}{2}\right) \\
&+
i\sin\left(\frac{\Delta t}{2}\right) \left\{\frac{W}{\Delta}+\frac{d\mathcal{E}_0\omega}{\omega^2-\Delta^2}\cos^2\left(\frac{\omega t}{2}\right)\right\}\biggr]. 
\end{split}
\end{equation}
At the end of the electric field region, $t=T\equiv L/v$ and $\omega T= 2\pi$ (regardless of $v$). Then, at the end of the region, the amplitude of state $|\psi^+_\uparrow\rangle$ is
\begin{equation}
c_+\left(T\right) = 2 e^{-i\Delta T/2}\sin\left(\frac{\Delta T}{2}\right) \left\{\frac{W}{\Delta}+\frac{d\mathcal{E}_0\omega}{\omega^2-\Delta^2}\right\}. 
\label{Cp2}
\end{equation}
Before the molecular beam exits the magnet, a laser beam depletes any remaining population in the odd parity $|\psi^-_\downarrow\rangle$ state by optically pumping to unobserved states via a short-lived even parity excited state, $|e^+\rangle$. Sec.~\ref{ParityEnergy} will explain why this step is necessary. In our experiment, typical values of the relevant parameters are
${d\mathcal{E}_0/(2\pi)\sim3.5}$ kHz, $\omega/(2\pi)\sim11.4$ kHz, and $\Delta/(2\pi) \sim 1-4$ kHz.

Ideally, we would detect the population of the even-parity state $|\psi^+_\uparrow\rangle$ immediately at this stage of the measurement. However, it is technically difficult to perform efficient state-selective detection inside the magnet. Instead, we allow the molecules to exit the magnet, and detect them outside. We refer to $|\psi^+_\uparrow\rangle$ ($|\psi^-_\downarrow\rangle$) as the detection (non-detection) state since we do (do not) measure its population. In the transition from inside to outside the magnet, the population in $|\psi^+_\uparrow\rangle$ is mapped, via adiabatic transport, onto the population of a particular resolved even-parity sublevel of the $|N^P = 0^+\rangle$ manifold of states. (More details on this mapping are discussed in Sec.~\ref{ParityEnergy}.) We detect the population of this sublevel using laser induced fluorescence. This yields a signal $S$, where 
\begin{eqnarray}
S &=& N_0 |c_+\left(T\right)|^2  \\
&\simeq& 4N_0\sin^2\left(\frac{\Delta T}{2}\right)\left[2\frac{W}{\Delta}\frac{d\mathcal{E}_0}{\omega}+\left(\frac{d\mathcal{E}_0}{\omega}\right)^2\right];
\label{signal}
\end{eqnarray}
here, $N_0$ is the number of molecules at $t=0$ in the odd-parity state $|\psi^-_\downarrow\rangle$, and we assumed $\Delta\ll\omega$ to simplify the expression. Note that the first term in the brackets is odd under $\mathcal{E}$-field reversal and the second term is even. Thus the parity violating asymmetry, $\mathcal{A}$, is extracted via the reversal of electric field $\mathcal{E}_0$:
\begin{equation}
\mathcal{A} = \frac{S\left(+\mathcal{E}_0\right)-S\left(-\mathcal{E}_0\right)}{S\left(+\mathcal{E}_0\right)+S\left(-\mathcal{E}_0\right)} = 2\frac{W}{\Delta}\frac{\omega}{d\mathcal{E}_0}. \label{asymmetry}
\end{equation}
Here, $\Delta, \mathcal{E}_0$ and $\omega$ are experimental values we control, and $d$ has been accurately measured previously \cite{CahnPRL}.
The uncertainty in the weak matrix element, $\delta W$, is limited to $\delta W\ge1/\left(2 \sqrt{2N_0}T\right)$ by shot noise (with minimum value obtained when $\Delta \ll 2/T$). This is exactly the shot-noise limited uncertainty expected for measurement of any energy shift, with $N_0$ detected particles observed for coherence time $T$. Hence, our technique can be interpreted as measuring an AC Stark shift between the pair of nearly-degenerate levels~\cite{Fortson93}.


\subsection{Subtleties of State Preparation, Depletion, and Detection}
\label{ParityEnergy}
At several steps in our experiment, laser-induced optical pumping is used to prepare and/or detect states.  The effect of these lasers, and the associated evolution of states before and after the lasers, proves rather subtle.  In this section, we explain the physics relevant to understanding these subtleties, with an emphasis on three critical concepts.

The first of these is that---due to the weak interaction, and to stray electric fields---the energy eigenstates in the system are not pure parity eigenstates. This distinction is unimportant outside the magnet, where the splitting between opposite-parity levels is ${\sim\!10^6}$ times larger than inside the magnet, and mixing of opposite-parity levels is negligible for our purposes. However, mixing of the relevant levels inside the magnet due to the weak interaction---though still small---is closely related to the effect we wish to measure, and cannot be neglected. We write the energy eigenstates inside the magnet as 
$|\psi^E_\uparrow(m_N,m_I)\rangle$ and $|\psi^E_\downarrow(m_N^\prime,m_I^\prime)\rangle$.These can of course be rewritten in the basis of parity eigenstates; in the absence of any stray electric field ($\mathcal{E}=0$), the Hamiltonian $H_\pm$ of Eqn.~\ref{Hplusminus} leads (via 1$^{\rm st}$ order perturbation theory) to 
\begin{eqnarray}
|\psi^E_\uparrow\rangle &\cong& |\psi^+_\uparrow\rangle + i(W/\Delta)|\psi^-_\downarrow\rangle, ~\mathrm{and} \nonumber \\
|\psi^E_\downarrow\rangle &\cong& |\psi^-_\downarrow\rangle +i(W/\Delta)|\psi^+_\uparrow\rangle.
\end{eqnarray}  

The second critical concept is that energy eigenstates inside and outside the magnet  map deterministically onto one another.  
Here, we mean specifically that energy levels with the same value of the conserved quantum number $m = m_s+m_N+m_I$ and the same nominal parity retain their ordering, independent of the strength of the magnetic field $\mathcal{B}$, and that changes in $\mathcal{B}$ are sufficiently slow as molecules move through the apparatus that all avoided crossings between such sublevels are fully adiabatic.  
In addition, $\mathcal{B}$ is large enough at all positions in the apparatus that $\delta m \neq 0$ (Majorana) transitions are deeply suppressed.  
Finally, everywhere in the apparatus except in the central interaction region where a large $\mathcal{E}$-field is deliberately applied, stray electric fields and the weak interaction are sufficiently small, and changes in $\mathcal{B}$ sufficiently rapid, that population transfer between states of nominally opposite parity is negligible even when such levels cross. 
Specifically, the energy eigenstates inside the magnet map onto energy (and parity) eigenstates outside the magnet that are associated with quantum numbers $N^P, J, F, m$, where we used standard definitions of the coupled angular momenta $\mathbf{J} = \mathbf{N} + \mathbf{S}$ and $\mathbf{F}=\mathbf{J}+\mathbf{I}$, and the total angular momentum projection $m$ is conserved as molecules move into and out of the magnet.  
In particular, $|\psi^E_\uparrow\rangle$ maps to $|N=0^+, J=1/2, F=1, m\rangle$, and $|\psi^E_\downarrow\rangle$  maps to $|N=1^-, J=1/2, F=1, m\rangle$.  
The mapping of superposition states preserves the relative population of the states, but scrambles their relative phase within the molecular ensemble (since minor changes in molecular beam position or velocity accumulate to phase differences  $\Delta \phi \gg 1$ over the long path into or out of the magnet.)  
For example, a superposition state $|\psi\rangle = \alpha |\psi^E_\uparrow\rangle + \beta |\psi^E_\downarrow\rangle$ inside the magnet maps to an incoherent mixture of $|N=0^+, J=1/2, F=1, m\rangle$ (with probabililty $|\alpha|^2$) 
and $|N=1^-, J=1/2, F=1, m\rangle$ (with probability  $|\beta|^2$), outside the magnet.

\begin{figure}
\includegraphics[width=83mm]{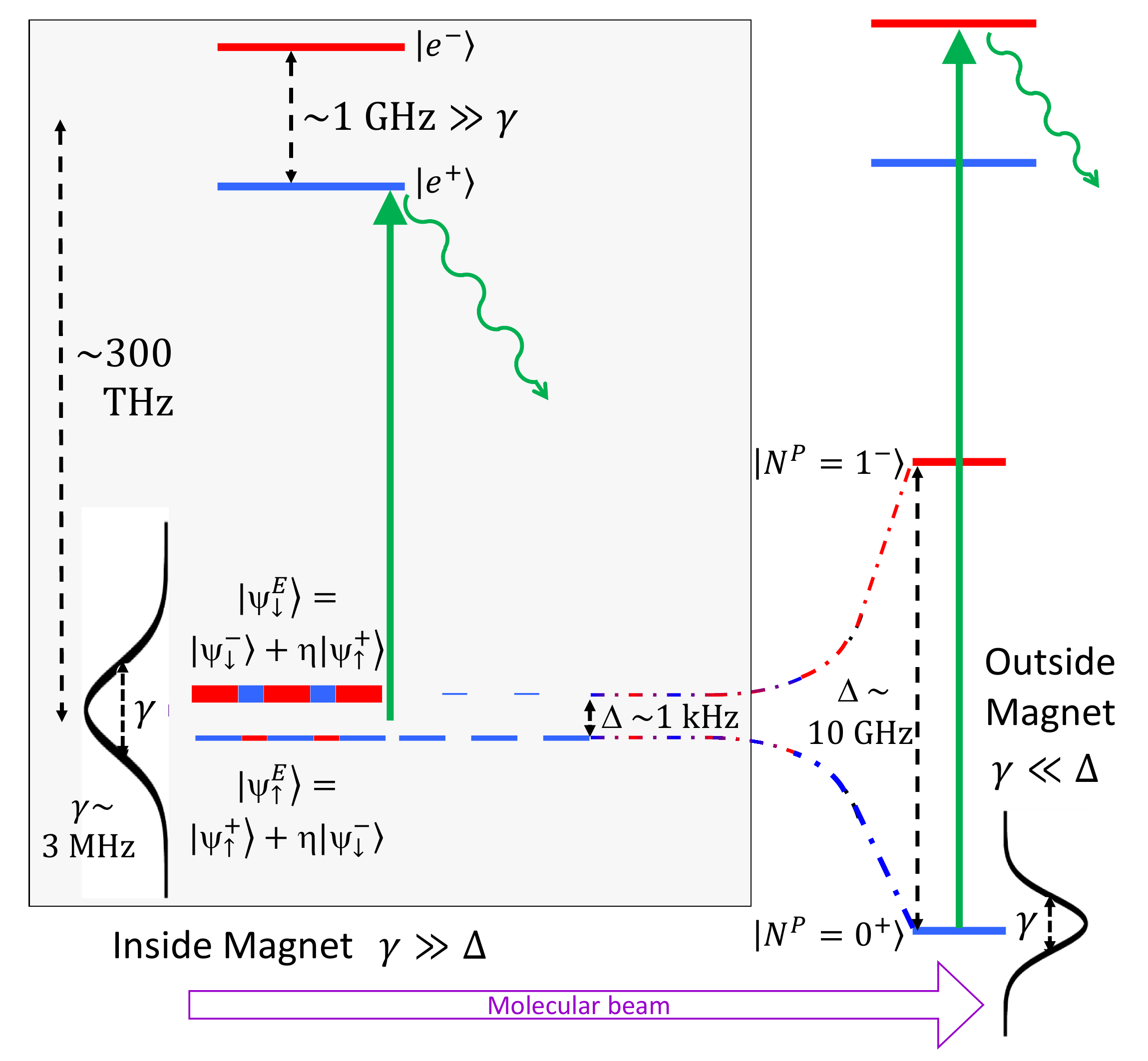}
\caption{(color online). Example of state projection and state evolution inside and outside the superconducting magnet. Shown here is the interaction with the 2$^{\rm nd}$ parity state projection laser inside the magnet, evolution of the resulting state as molecules leave the magnet, and interaction with the detection laser outside the magnet. Even (odd) parity states are shown in blue (red). The natural linewidth of the optical transitions is $\gamma/(2\pi)\sim3$ MHz. Inside the magnet, energy eigenstates in the ground state ($|\psi^{\rm E}_\uparrow\rangle$ and $|\psi^{\rm E}_\downarrow\rangle$) are separated in energy by $\Delta/(2\pi)\sim1$ kHz $ \ll \gamma$, so these states are fully unresolved by the optical transition.   The state projection laser addresses an excited state of definite parity, and hence excites the particular superposition of ground states that is a parity eigenstate.  Here, the excited state of the transition, $|e^+\rangle$, has even parity, so the laser depletes the odd parity ground state $|\psi_\downarrow^-\rangle$ and leaves behind the superposition corresponding to the even parity state $|\psi_\uparrow^+\rangle$.  The population of each energy eigenstate is adiabatically transported to a corresponding state $|N^P \rangle$ (a simultaneous eigenstate of energy and parity) outside the magnet, according to the mapping $|\psi^{\rm E}_\uparrow\rangle \leftrightarrow |N^P = 0^+\rangle$ and $|\psi^{\rm E}_\downarrow\rangle \leftrightarrow |N^P =1^-\rangle$. Here, outside the magnet, these eigenstates are separated in energy by $\Delta \sim 2\pi\times 10$ GHz $\gg \gamma$, and hence are fully resolved by the optical transition.  Tuning the detection laser to the transition between $|N^P = 0^+\rangle$  and a different excited state of definite parity, $|e^-\rangle$, results in a fluorescence signal proportional to the population of the $|N^P = 0^+\rangle$ state, and hence of the  $|\psi^{\rm E}_\uparrow\rangle$ state that had been in the magnet after interaction with the state projection laser. As described in the text, this population is, to an excellent approximation, the same as the population $|c_+|^2$ of the even parity state $|\psi_\uparrow^+\rangle$ \textit{prior} to interaction with the state projection laser.  This is exactly the signal desired for our NSD-PV measurement.}
\label{fig:InsideMagnerParityLevels}
\end{figure}

The third critical concept is whether, when the laser excites molecules, the energy eigenstates of interest are well-resolved or completely unresolved.  These refer, respectively, to cases where the splitting between these states is large or small compared to the natural linewidth of the optical transition, $\gamma \approx 2\pi\!\times\! 3$ MHz (based on the measured lifetime~\cite{Berg89} of the state $A ^2\Pi_{1/2}$ excited by the lasers).  Outside the magnet, where the levels are well-resolved, optical pumping by the laser depletes an energy eigenstate (which is, again, also a parity eigenstate), and the emitted fluorescence intensity is proportional to the population of that eigenstate.  By contrast, inside the magnet---where the levels are unresolved---interaction with the laser depletes the particular superposition of energy eigenstates that corresponds to a parity eigenstate.  This is because the excited state of the optical transition is an isolated level with definite parity. (The $\mathcal{B}$-field does not have the correct value to induce near-degeneracies of opposite-parity levels in the $A ^2\Pi_{1/2}$ excited state.) Hence, inside the magnet, the laser acts to project the wavefunction (or more generally, the density matrix) of the system onto a parity eigenstate.  The parity quantum number ($\pm 1$) of the state projected onto is the same as that of the excited state addressed by the laser.

Now we can understand more precisely the effect of the lasers at each step in Fig.~\ref{fig:ApparatusSchematic}.  
In step 3, the 1$^{\rm st}$ parity state projection laser, inside the magnet, is tuned to resonance with the odd-parity excited state $|e^-\rangle$, and so projects the initial incoherent superposition of $|\psi^+_\uparrow\rangle$ and $|\psi^-_\downarrow\rangle$ onto the pure parity eigenstate $|\psi^-_\downarrow\rangle$.  
This is exactly the initial state preparation described in Sec.~\ref{IdealMeasure}.  
Then, following application of the electric field $\mathcal{E}(t)$ (step 4 in Fig.~\ref{fig:ApparatusSchematic}), the system is in the superposition state $|\psi(T)\rangle \cong |\psi^-_\downarrow\rangle + c_+(T)|\psi^+_\uparrow\rangle$, where $c_+(T)$ is given by Eqn.~\ref{Cp2}.  
Next, interaction with the 2$^{\rm nd}$ parity state projection laser, tuned to resonance with the even parity excited state $|e^+\rangle$, projects the system onto $|\psi^+_\uparrow\rangle$ (step 5 in Fig.~\ref{fig:ApparatusSchematic}).  
Hence, after interacting with this laser for a short time $T_b$, the system is in the (unnormalized) state $|\psi(T+T_b)\rangle = c_+(T)|\psi^+_\uparrow\rangle$. 
To understand the evolution of this state as it leaves the magnet, we write it in the basis of energy eigenstates: 
$|\psi(T+T_b)\rangle \cong c_+(T)|\psi^E_\uparrow\rangle - ic_+(T)(W/\Delta)|\psi^E_\downarrow\rangle$. This maps onto energy eigenstates outside the magnet as described above, with the probability to be in state $|N=0^+, J=1/2, F=1, m\rangle$ given by $ |c_+(T)|^2$. The detection laser excites this resolved state (step 6 in Fig.~\ref{fig:ApparatusSchematic}), and we observe a fluorescence signal proportional to this probability. This is exactly the signal $S$ described above in Sec.~\ref{IdealMeasure}. 

This understanding also clarifies the importance of the 2$^{\rm nd}$ parity state projection laser in our experimental protocol. Suppose this laser were absent.  Then, just after the electric field ends at time $T$, the state of the system can be written in terms of energy eigenstates, with the result 
\begin{equation}
|\psi(T)\rangle \cong [c_+(T)-\frac{iW}{\Delta}]|\psi^E_\uparrow\rangle + [1-c_+(T)-\frac{iW}{\Delta}c_+(T)]|\psi^E_\downarrow\rangle.
\end{equation}
Then, transport out of the magnet and detection of the $|N=0^+, J=1/2, F=1, m\rangle$ state would result in a signal $S \propto |c_+(T)-i(W/\Delta)|^2$ rather than $S\propto |c_+(T)|^2$ as desired. The resulting NSD-PV asymmetry would be roughly a factor of 2 smaller in this undesired case, compared to the case we actually employ. In addition, the 2$^{\rm nd}$ parity state projection laser dramatically reduces the sensitivity of the signal to stray $\mathcal{E}$-fields in the region inside the magnet (where the levels remain close in energy and can be significantly mixed by such small fields), but after the deliberately-applied field $\mathcal{E}(t)$ has ended~\cite{AltuntasThesis}. In fact, for diagnostics of such electric fields, we deliberately block the 2$^{\rm nd}$ parity state projection laser beam. Similarly, for diagnostics of stray $\mathcal{E}$-fields at locations prior to application of $\mathcal{E}(t)$, we replace the 1$^{\rm st}$ parity state projection laser with a projection laser outside the magnet that depletes the state $|N=0^+, J=1/2, F=1, m\rangle$, which then maps onto $|\psi^+_\uparrow\rangle$ inside the magnet (though this auxiliary laser is not shown in Fig.~\ref{fig:ApparatusSchematic}.).

\subsection{Measurement Principle: Modifications to Account for Experimental Constraints}
\label{PrincipleExp}

\subsubsection{Realistic Experimental Geometry}
The expression for the signal given in Eqn.~\ref{asymmetry} must be modified
to account for an experimental geometry that differs somewhat from the ideal case described in Sec.~\ref{IdealMeasure}. As before, we have the initial conditions $c_+\left(0\right) =0$ and $c_-\left(0\right) =1$ due to the action of the 1$^{\rm st}$ parity state projection laser. However, in our apparatus there is a spatial (and hence also temporal) gap between the application of the parity state projection lasers and the start/end of the $\mathcal{E}$-field (see Sec.~\ref{IR}). We define free evolution times $T_{f1}$ and $T_{f2}$ such that the 1$^{\rm st}$/2$^{\rm nd}$ parity state projection laser ($\rm{L}_{P1}$/$\rm{L}_{P2}$) is applied with time difference $T_{f1} / T_{f2}$ before/after the Stark interference $\mathcal{E}$-field start/end. Accounting for these additional free evolution times, the applied Stark interference $\mathcal{E}$-field is defined as
\begin{equation}
\mathcal{E}\left(t\right) =
\begin{cases}
\mathcal{E}_0 \sin\left[\omega \left(t - T_{f1}\right)\right],& \text{for } T_{f1}< t < T_{f1}+T_e\\        0,         & \text{otherwise}.
\end{cases}
\label{Efield}
\end{equation}
Here, $T_e = 2\pi/\omega$ is the time duration of the $\mathcal{E}$-field pulse. We integrate the Schr{\" o}dinger equation from $t=0$ to $t=T=T_{f1}+T_e+T_{f2}$ and find 
\begin{eqnarray}
c_+\left(T\right) &&= \frac{iW}{\Delta}\left[e^{-i\Delta T}-1\right] \nonumber \\ &&+  \frac{2 d\mathcal{E}_0 \omega}{\omega^2 -\Delta^2}e^{-i\Delta \left(T_e/2+T_{f1}\right)}\sin\left[\frac{\Delta T_e}{2}\right].
\end{eqnarray}
At time $T$, the state of the system is $| \psi\left(T\right)\rangle = c_+(T)|\psi_{\uparrow}^{+}\rangle + e^{-i\Delta T} |\psi_{\downarrow}^{-}\rangle$. Then, as described in the previous section, the action of the second parity state projection laser $\rm{L}_{P2}$, transport out of the magnet, and detection together lead to the measured signal ${S= N_0 |c_+\left(T\right)|^2}$, given by
\begin{equation}
\begin{split}
S & \simeq  4N_0\left(\frac{ d\mathcal{E}_0 \omega}{\omega^2 -\Delta^2}\right)\left\{\frac{ d\mathcal{E}_0 \omega}{\omega^2 -\Delta^2} \sin^2\left[\frac{\Delta T_e}{2}\right] \right.\\
&\left.+2\frac{W}{\Delta} \sin\left[\frac{\Delta T_e}{2}\right] \sin\left[\frac{\Delta}{2}{ T}\right]\cos\left[\frac{\Delta}{2}{\left(T_{f1}-T_{f2}\right)}\right]\right\}.
\end{split}
\end{equation}
As the last step, we compute NSD-PV asymmetry, $\mathcal{A}_{thy}$, associated with reversal of $\mathcal{E}_0$, following Eqn.~\ref{asymmetry}: 
\begin{eqnarray}
\mathcal{A}_{thy} &=& 2\frac{W}{\Delta}\frac{\omega^2 -\Delta^2}{ d\mathcal{E}_0 \omega}
\frac{\sin\left[\frac{\Delta}{2}{ \left(T_e+T_{f1}+T_{f2}\right)}\right]}{\sin\left[\frac{\Delta}{2}{ T_e}\right]} \nonumber\\
&& \times \cos\left[\frac{\Delta}{2}{\left(T_{f1}-T_{f2}\right)}\right].
\label{AsymParity}
\end{eqnarray}
Here, as before, $\Delta$ and $\mathcal{E}_0$ are experimental values we control, the parameters $\omega,T_e, T_{f1}$, and $T_{f2}$ are defined by the geometry of the interaction region (discussed later), the molecular velocity $v$ is measured to be $v = 616$ m/s, and the dipole matrix element at each crossing, $d$, has been measured previously~\cite{CahnPRL}.

\subsubsection{Imperfect Optical Pumping}
\label{OPR}

\begin{table}[h!]
	\begin{tabular}{ l  l l }
		\hline
		\hline
		Name & $\mathcal{E}$-field & $\rm{L}_{P1}$ and $\rm{L}_{P2}$ Lasers Shutter State\\
		\hline	
		$S(-\mathcal{E}, o)$, & $-\mathcal{E}_r$ & Open \\
		$S(+\mathcal{E}, o)$, & $+\mathcal{E}_r$ & Open \\
		$S(0, o)$, & None & Open \\
		$S(0, c)$, & None & Closed \\
		$S(\pi, o)$, & $\pi$-pulse & Open \\
		\hline
		\hline 	 
	\end{tabular}
	\caption{$\mathcal{E}$-field conditions and shutter states for the five signals employed in a NSD-PV measurement.}
	\label{TableSignals}
\end{table}

In our discussion up to now, we assumed perfect depletion by both parity state projection lasers $\rm{L}_{P1}$ and $\rm{L}_{P2}$; however, in our experiment we observe incomplete depletion (possibly due to inadequate laser power and/or interaction time, or from repopulation of the depleted states by decay from metastable states populated in the molecular beam source). In order to account for these effects, we record several types of signals, listed in Table~\ref{TableSignals}. We define signals $S(\mathcal{E}, s)$, where $\mathcal{E}$ indicates the $\mathcal{E}$-field condition and $s$ is the state of shutters that are used to block both lasers $\rm{L}_{P1}$ and $\rm{L}_{P2}$ simultaneously. The $\mathcal{E}$-field condition takes several possible values, denoted as $\pm\mathcal{E}_r$ (corresponding to the reversing $\mathcal{E}$-field of Eqn.~\ref{AsymParity}, with amplitude $\mathcal{E}_0$ positive or negative, respectively); 0 (corresponding to no applied $\mathcal{E}$-field); and $\pi$ (corresponding to a unipolar $\mathcal{E}$-field pulse with magnitude and duration set to achieve complete population swapping between the states $|\psi^+_\uparrow\rangle$ and $|\psi^-_\downarrow\rangle$, discussed further below). The shutter condition is assigned values $s = o$ [open] and $s=c$ [closed], corresponding to all state projection laser light on and off, respectively~\cite{CahnPRL}.  

The depletion efficiency for the state projection lasers is related to certain ratios of these signals, which we define as the ``optical pumping ratios" (OPRs), $R_{i}(\mathcal{E}) = \frac{S(\mathcal{E},o)}{S(\mathcal{E},c)}$.
The subscript $i$ indicates the parity state projection laser $L_{\rm P1}$ or $L_{\rm P2}$. For example, the depletion efficiency for $\rm{L}_{P1}$ is $1-R_1(0)=1-\frac{S(0,o)}{S(0,c)}$, where for perfect depletion $R_{1}(0)=0$. Recall that $\rm{L}_{P2}$ depletes the non-detection level, the odd parity ground state $|\psi^-_\downarrow\rangle$. With our current setup, we cannot measure the remaining population in this level directly via laser-induced fluorescence. Instead, we apply a $\pi$-pulse after laser $\rm{L}_{P2}$ so that the population in this non-detection state is transferred to the detection state, and vice-versa. The resultant signal, $S(\pi, o)$, is taken with both state projection lasers applied. Thus, the depletion efficiency for the 2$^{\rm nd}$ projection laser is $1-R_{2}(\pi)$. Typically we have $R_{1}(0)\approx\!3-6 \%$ and $R_{2}(\pi)\approx\!7-10\%$. 

To measure the fraction of the (odd-parity) non-detection state population that is transferred to the (even-parity) detection state by the Stark interference $\mathcal{E}$-field pulse and NSD-PV effects, despite a nonzero value of $R_{1}(0)$ (i.e. imperfect depletion), we define the state transfer efficiency (STE) as
\begin{equation}
\rm{STE}(\mathcal{E})  = \frac{R_{1}( \mathcal{E}) - R_{1}(0)}{1- R_{1}(0)}= \frac{S( \mathcal{E},o) - S(0,o)}{S(0,c) - S(0,o)}. 
\label{STE1}
\end{equation}
Following Eqn.~\ref{asymmetry}, the NSD-PV asymmetry in terms of the measured signals is
\begin{eqnarray}
\mathcal{A} &=& \frac{\rm{STE}(+\mathcal{E}) -\rm{STE}(-\mathcal{E}) }{\rm{STE}(+\mathcal{E}) +\rm{STE}(-\mathcal{E})}\nonumber\\
&=&\frac{S( +\mathcal{E},o) -S( -\mathcal{E},o) }{S( +\mathcal{E},o) + S( -\mathcal{E},o)-2S(0,o)}.
\end{eqnarray}

\section{Apparatus}

\subsection{Molecular Beam Source}
We create a beam of BaF using a supersonic free jet source based on Ref.~\cite{Hinds}. A mixture of argon ($95\%$) and $\mathrm{SF}_6$ ($5\%$) at pressure ${\sim\!13}$ atm expands into the vacuum chamber through a pulsed valve nozzle with $1$ mm diameter. The valve is driven by an electrical pulse of $280$ $\mu$s duration. A target, consisting of a strip of Ba metal affixed to the outer diameter of a 2 mm thick, 10 cm diameter wheel, located just under the jet aperture, is ablated by a pulsed Nd:YAG laser, with typical pulse energy ${\sim\!20}$ mJ and pulse duration ${\sim\! 8}$ ns, fired ${\sim\!340}~\mu s$ after the start of the electrical pulse. A plume of Ba is entrained in the jet and BaF molecules form through collisions between Ba and $\mathrm{SF}_6$. The rapid adiabatic expansion of the gas into the vacuum chamber cools the molecules' rotational energy and longitudinal velocity to temperature $T{\sim30}$ K. The beam has a forward mean velocity $\bar{v}=616$ m/s with FWHM spread $\delta v/\bar{v}=7\%$; each pulse of BaF has typical duration ${\sim\! 100~\mu}$s at the beam source and ${\sim\! 500 ~\mu}$s in the detection region. We create molecular beam pulses at  $10$ Hz repetition rate.

The source chamber is separated from the rest of the apparatus by a skimmer with $6.35$ mm diameter, located $6$ cm from the source. This allows differential pumping between the source chamber ($\sim\!10^{-5}$ Torr) and the rest of the experiment ($10^{-7} - 10^{-8}$ Torr). After the skimmer, the molecules enter a second vacuum chamber where a laser interaction region is used to probe the molecules for beam diagnostics and/or to prepare the molecules in a specific state (via optical pumping) for certain auxiliary measurements. Collimators define the molecular beam to have radius $R_b \approx 3.8$ mm in the center of the main interaction region.

\subsection{Magnetic Field Control}
A superconducting (SC) magnet with a room-temperature bore of ${\sim\!22}$ cm diameter generates the $\mathcal{B}$-field. The commercial magnet package has $5$ SC and $14$ room temperature (RT) gradient coils for adjusting the $\mathcal{B}$-field homogeneity. One of the RT shim coils (the Z0 coil, which provides a uniform field) is employed to tune the $\mathcal{B}$-field (and thus the detuning $\Delta$ between the two levels of opposite parity). An additional set of shim coils, referred to as the  \textquotedblleft mini-shims", are used to create local variations in the $\mathcal{B}$-field and hence to shape the field for maximum homogeneity and/or to generate field gradients of various shapes for systematic error tests. These home-built coils are wound around the molecular beam line, with radius 2 cm and spaced 1 cm apart.  Each mini-shim coil produces a magnetic field with FWHM length of 6 cm along $z$, and an amplitude of up to 120 mGauss on axis.

Our measurement technique requires exceptional magnetic field homogeneity over the region between the parity state projection lasers. The homogeneity is measured and improved in a multi-step process. An array of $32$ NMR probes, spatially distributed around the magnet center, is used for initial $\mathcal{B}$-field measurement with precision $\delta \mathcal{B} / \mathcal{B} \sim 0.5$ ppm \cite{Murphree2007160}. (One of these probes is used throughout the measurements to measure the field tuning achieved using the Z0 coil, and to account for small overall drifts in the $\mathcal{B}$-field.) When the value of $\mathcal{B}$ is initially tuned to any given level crossing, $\mathcal{B}$-field homogeneity is first optimized by adjusting the SC and RT shim coils to minimize the r.m.s. deviation among the NMR probe readings. For finer precision, we use signals from BaF molecules to measure the field, and the RT and mini-shim coils to correct residual inhomogeneities (see Sec.~\ref{BfieldMeasure}).

\subsection{Electric Field Control}
\label{IR}
A complex interaction region (IR) enables control over $\mathcal{E}$-fields in the system and, simultaneously, delivery of laser light to the molecular beam inside the magnet (where we have no physical access once under vacuum). The IR is designed to minimize $\mathcal{B}$-field inhomogeneities due to the small, but non-negligible, magnetic susceptibility of its parts. We use a simple principle to minimize $\mathcal{B}$-field inhomogeneities due to the material of the IR. That is: if a long, straight object with constant cross-section is uniformly magnetized along its length, the field due to the magnetization is like that of an infinite solenoid: zero outside the object, and uniform inside. Hence, to the extent possible we construct the IR from materials with small linear susceptibility (which will be uniformly magnetized along the uniform $\boldsymbol{\mathcal{B}}$-field) and with constant cross-section along the $\boldsymbol{\mathcal{B}}$-field axis.

\begin{figure*} 
\includegraphics[width=180mm]{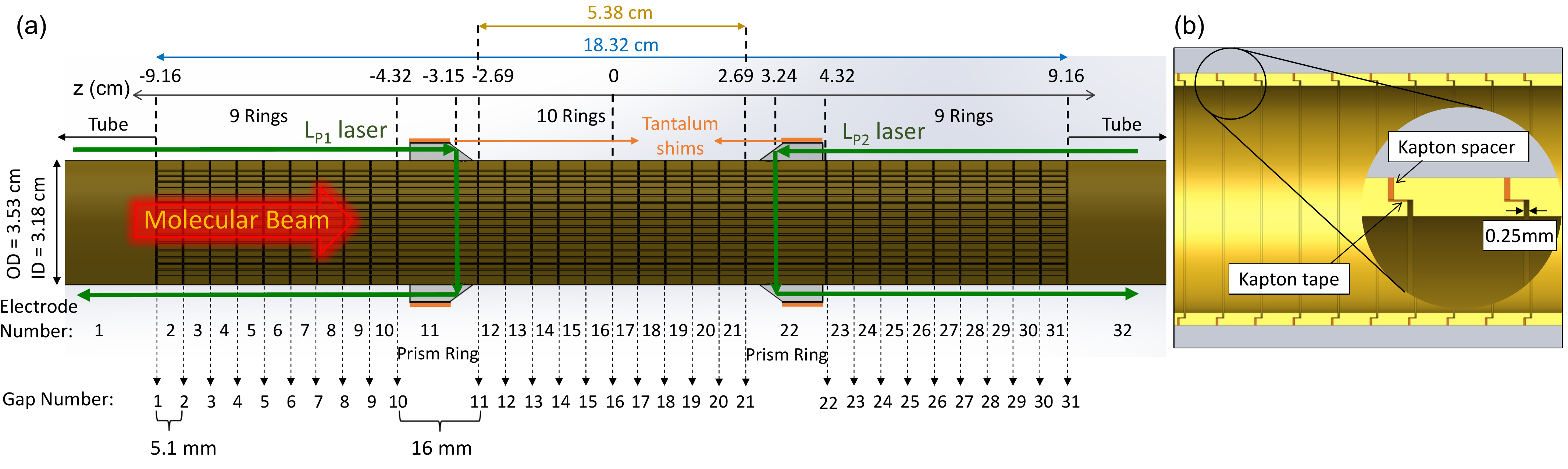}
\caption{ (color online) (a) Schematic of the assembled IR and the relevant coordinates. The IR has 32 cylindrical electrodes and 31 gaps between electrodes. The central 10 rings (numbered from 12 to 21, corresponding to the region between gaps 11 and 21) encompass the region where reversing the sine wave $\mathcal{E}$-field is nonzero during the NSD-PV measurements. Thin arrows (green) signify the parity state projection laser beam paths, including their reflections in glass prisms. Paramagnetic tantalum foil shims (orange) are attached to the prisms to minimize the $\mathcal{B}$-field nonuniformity induced by the diamagnetic glass. The axial lines on electrodes 2-31 are grooves that act as wire channels to prevent electrode connection wires from twisting and causing $\mathcal{B}$-field inhomogeneities. The wire channels also provide strain relief at the wire solder points. (b) Cross section depicting the placement of Kapton tape and spacer insulating (respectively) the radially- and axially-oriented connection surfaces between rings. The insulated surfaces have no line-of-sight to the molecules, so static charges that may build up on these surfaces are of no concern.}
\label{fig:IRCoordinates}
\end{figure*}

The IR (shown in Fig.~\ref{fig:IRCoordinates}) has 32 co-axial cylindrical electrodes: 2 tubes (used as endcaps), 28 rings, and 2 extra-wide rings, all machined from OFHC copper with outer diameter $=3.53$ cm and inner diameter $=3.18$ cm. Standard rings each have a length of $6$ mm; the extra-wide rings, referred to as prism rings, are ${\sim\!17}$ mm long to mount prisms that reflect laser light through holes in the rings. Two electrical connections are made to each electrode: one to apply a voltage and other to measure the applied voltage. To ensure a clean and uniform surface, the IR electrodes were bright-dipped and then plated with high-purity gold (with a palladium underlayer as a diffusion barrier). We maintain concentricity between stacked electrodes using an interlocking male-female design. Axially oriented surfaces between each electrode are insulated using laser-machined Kapton spacers, and radially oriented surfaces have Kapton tape for insulation. The Kapton thickness ($0.25$ mm) was minimized to maintain $\mathcal{B}$ and $\mathcal{E}$-field uniformity across the IR. 

For PV data, we apply a complex set of voltages to the central 10 rings that lie between the prism rings to generate the sharp turn-on and off of a pure sine wave. The 9 rings that lie outside the prism rings on either side allow us to measure and control the $\mathcal{E}$-field in locations outside the region between the parity state projection lasers. We use this system to deliberately apply $\mathcal{E}$-fields of various shapes, in order shim out inhomogeneities due to stray fields, and/or to study  possible systematics. Each electrode is controlled with an individual digital-to-analog voltage channel with minimum resolution of $0.9$ mV, corresponding to a maximum change in the $\mathcal{E}$-field on the central axis of the molecular beam of ${\sim\!0.3}$ mV/cm.

\subsection{Lasers and Optics}

Four external cavity diode lasers (ECDLs) address the transitions for parity state projection and detection. A frequency-stabilized HeNe laser is employed as a frequency reference, and is used in conjuction with a Fabry-P\'{e}rot transfer cavity to stabilize the ECDL frequencies, with a system similar to that in Ref.~\cite{Zhao98}. Three of the ECDLs are tuned to different transitions within the $X ^2\Sigma^+(v=0) \rightarrow A ^2\Pi_{1/2}(v^\prime = 0)$ manifold of transitions, at wavelength $\lambda \approx 860$ nm. The two ECDLs used as the two parity state projection lasers are tuned to different frequencies when the magnetic field is tuned to a different level crossing. 
The third ECDL is used both to excite the first step of a two-photon transition used for detection (see below), and also to deplete the molecular state outside the magnet for auxiliary measurements. The fourth ECDL, used only for detection, is tuned to a rotational line of the $A ^2\Pi_{1/2}(v^\prime = 0)\rightarrow D ^2\Sigma^+(v=0)$ transition, at wavelength $\lambda \approx 797$ nm. 

An optical mount, consisting of beam shaping optics and beam steering mechanics, is attached to each end of the IR to align laser beams through the molecular beam at the desired locations. A beam-shaping telescope expands the laser light to a vertical $1/e^2$ diameter of $13.2$ ($11.5$) mm for the first (second) parity state projection laser. This corresponds to a laser beam intensity  at the edges of the molecular beam that is about $50 \%$ of the peak intensity. Following the telescope, the laser beam enters a prism attached to a prism ring on the IR and is reflected by 90 degrees so that it traverses the molecular beam. Then, the laser beam reflects by 90 degrees again through a second prism, so that it exits the IR and strikes a photodiode, which is used to verify transmission of the laser beam through the apparatus. The prisms are BK7-glass right-angle prisms cut to a width of $4.8$ mm along the molecular beam direction. The prism surface exposed to the molecules is coated with conductive indium tin oxide, and electrically connected to the prism ring with a thin layer of indium metal, to ensure that no static charges build up on the inside surface. Additionally, we attached paramagnetic tantalum foil shims to the prisms in order to minimize the $\mathcal{B}$-field nonuniformity induced by the diamagnetic BK7-glass. We optimized the tantalum shim thickness such that the calculated peak-to-valley $\mathcal{B}$-field variation experienced by the molecules is minimized; we find $\delta B/B \lesssim 0.03$ ppm. 

\subsection{Detection} 

We measure the population of the detection state after the molecules exit the magnet, via laser-induced fluorescence with a photomultiplier tube (PMT). A custom large solid angle focusing mirror pair, based on the design in Ref.~\cite{Shimizu83}, is employed to maximize the light collection efficiency. A 1 inch diameter fused silica light guide transports light from the focal point of the mirrors, in vacuum, to the PMT outside. With this arrangement, $\approx\! 85\%$ of the fluorescence light is directed into the PMT. 

We employ a two-photon excitation scheme to realize low-background fluorescence detection. A first laser beam is applied to excite the $X{^2}\Sigma^{+}\rightarrow A^{2}\Pi_{1/2}$ transition, and a second laser beam, spatially overlapped with the first, is tuned to drive the $A^{2}\Pi_{1/2} \rightarrow D{^2}\Sigma^{+} $ transition. We detect fluorescence from the $D{^2}\Sigma^{+} \rightarrow X^{2}\Sigma^{+}$ decay, at $413$ nm. This scheme makes it easy to reduce background from scattered laser light, since the detection wavelength is far to the blue of both laser wavelengths. A red-blocking, blue-transmitting (BG40) color glass filter is located at the PMT entrance for this purpose. 
 

\section{System Characterization}
\label{SystemCharacterization}

In this section, we describe a set of measurements used to optimize and characterize the electric and magnetic fields experienced by molecules in the course of our NSD-PV measurements. Measurements of both $\mathcal{E}$- and $\mathcal{B}$-fields rely on applying local pulses of $\mathcal{E}$-field, so we begin by describing how these pulses are created. Then we describe specifics of the $\mathcal{B}$- and $\mathcal{E}$-field measurements.


\subsection{Applying Local $\mathcal{E}$-Field Pulses}
\label{localPulses}

The IR electrodes are used to control the $\mathcal{E}$-field within the central region of the magnet, as follows. Consider the case where all rings before the gap location $z_k$, where $k$ indicates the gap number between two electrodes, are set at one voltage, $V_0$, and all rings after gap $k$ are set at $V_0 + \delta V$. The voltage step gives rise to an electric field which, near the axis of the tube, is well-approximated by 
\begin{equation}
\mathcal{E}^u(z ;z_k) ={\mathcal{E}}_{0}^{u}  \text{sech} \left(\frac{z - z_k}{\sigma_u} \right),
\label{UniEr}
\end{equation} 
where $\sigma_u=0.76$ cm and $\mathcal{E}_{0}^{u}=0.42~\delta \rm V$/cm. We refer to this as a unipolar $\mathcal{E}$-field pulse. Over the range of axial positions $0< \rho < 3.8$ mm, where detected molecules are present in the IR, this $\mathcal{E}$-field varies by $<3\%$; going forward we ignore this small variation and treat the $\mathcal{E}$-field as radially uniform.  

\begin{figure}
	\includegraphics[width=83mm]{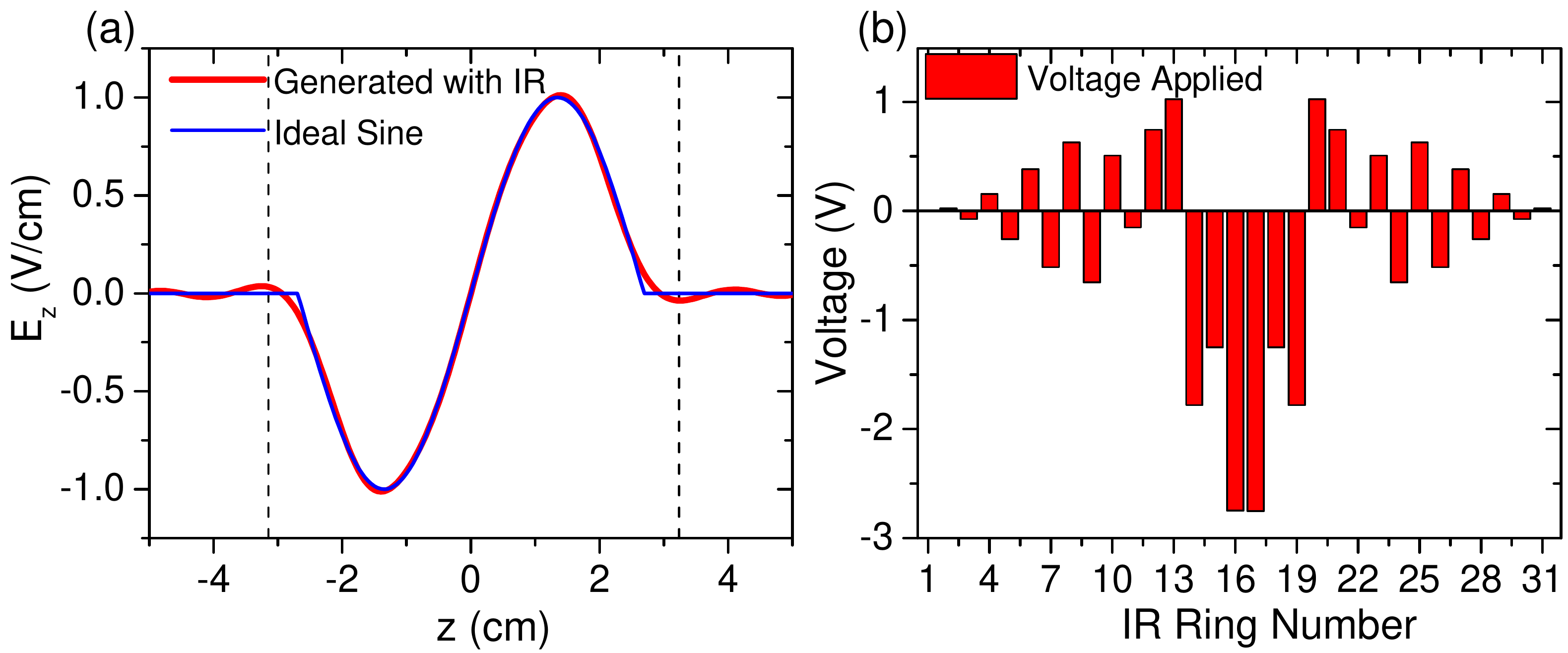}
	\caption{(color online) (a) Calculated shape of our experimentally-applied $\mathcal{E}$-field. Also shown is the desired analytic form of the pulse. The calculated field shape is shown on the molecular beam axis. Vertical dashed lines indicate the location of the parity state projection laser beams. (b) Voltages applied to the IR ring electrodes to generate the field shown in (a).}
	\label{fig:UnipolarEfield}
\end{figure}

The width in $z$ of the unipolar $\mathcal{E}$-field pulse, $\sigma_u$, is much smaller than the usual spacing between ring gaps, $z_{k+1}-z_{k}$. Hence it is possible to create nearly any desired shape of $\mathcal{E}(z)$ by properly superposing fields from voltage steps at each gap. For example, this principle is used to create the finite-width sinusoidal pulse described in Sec.~\ref{PrincipleExp} (see Fig.~\ref{fig:UnipolarEfield}). In addition, for various diagnostic tests (described below) we frequently apply unipolar $\mathcal{E}$-field pulses at different positions in the IR. We also sometimes use a bipolar pulse, which has two extrema and is generated by applying equal and opposite voltage steps $\pm \delta V$ at adjacent ring gaps, i.e., a voltage offset $\delta V$ to a single ring. The field resulting from such a voltage configuration, $\mathcal{E}_b(z; z_r)$, is well-approximated by the functional form  
\begin{equation}
\mathcal{E}^{b}(z; z_r) = {\mathcal{E}}_{0}^{b}  \text{sech} \left(\frac{z - z_r}{\sigma_b} \right)\text{tanh} \left(\frac{z - z_r}{\sigma_b} \right),
\label{BiEr}
\end{equation} 
where $\sigma_b =0.82$ cm, ${\mathcal{E}}_{0}^{b}(\delta \rm V) =0.28~\delta \rm V$/cm, and $z_r$ is the center of the ring where the voltage offset is applied.

The spatially narrow unipolar $\mathcal{E}$-field pulse induces a population transfer between the nearly-degenerate opposite-parity states~\cite{CahnPRL}. Following Eqn.~\ref{Equ:Shro} (with $W = 0$), assuming as usual initial conditions $c_+(0) = 0$ and $c_-(0) = 1$, and in the limit that the amplitude $c_+ \ll 1$ (where $1^{st}$ order perturbation theory holds), the amplitude of the even parity state is determined by the relation
\begin{equation}
\dot{c}_+\left(t\right) = -id\mathcal{E}(t)e^{-i\Delta t}.
\label{Equ:evenState1stOrder2}
\end{equation}
Here we take $t=(z-z_{\rm L_{P1}})/v$, with molecular beam speed $v$ and $z_{\rm L_{P1}} = -3.15$ cm the position of the 1$^{\rm st}$ parity state projection laser. We integrate the above equation and find the transfer amplitude,
\begin{equation}
c_+\left(T\right) = -id \int_{0}^{T}\mathcal{E}(t)e^{-i\Delta t}dt.
\label{Equ:EnrC+}
\end{equation}
Assuming that no population transfer can occur outside the range, $0<t<T$, we can extend the integration limits to infinity:
\begin{equation}
c_+(T)=-id\int_{-\infty}^{+\infty}\mathcal{E}\left(t\right)e^{-i\Delta t} dt = -id\sqrt{2\pi}\tilde{\mathcal{E}}\left(\Delta \right),
\label{Equ:Enr2} 
\end{equation} 
where $\tilde{\mathcal{E}}\left(\Delta\right)$ is the Fourier transform of ${\mathcal{E}}\left(t\right)$~\cite{CahnPRL}.
Then the resultant signal, $S$, is given by
\begin{equation}
S \propto \lvert c_+(T) \rvert ^2 \propto |\tilde{\mathcal{E}}(\Delta)|^2, 
\label{eqn:signalEPulse}
\end{equation}
where $
\tilde{\mathcal{E}}\left(\Delta\right)=\sqrt{{\pi}/{2}}\sigma_t\mathcal{E}_{0}^u \text{sech}\left(\pi \Delta\sigma_t/2\right)$, with $\sigma_t=\sigma_{u}/v=12.3~\mu$s.


\subsection{Local Magnetic Field Measurement}
\label{BfieldMeasure}

\begin{figure}
	\includegraphics[width=83mm]{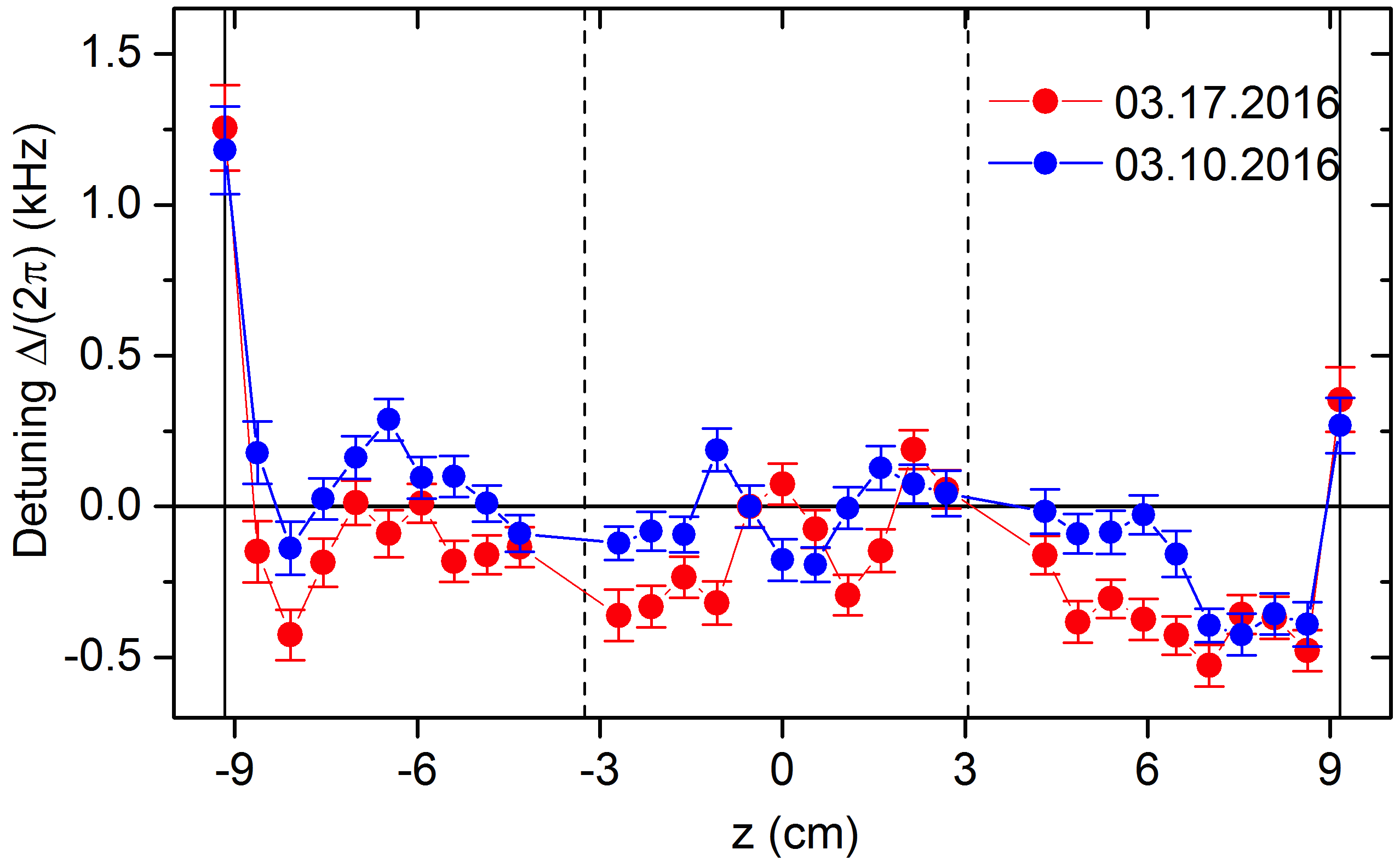}
	\caption{(color online). Typical magnetic field homogeneity after shimming. Vertical dashed lines indicate parity state projection laser locations, and solid lines denote the region of the IR where electrodes control the $\mathcal{E}$-field. $\Delta = 0$ was chosen to facilitate visibility of the fluctuations. Measurements were made with the same currents applied to all shim and mini-shim coils, one week apart (dates indicated in legend). Both measurements have $\delta \mathcal{B}/\mathcal{B}<0.02$ ppm (r.m.s.) in the region between $\rm{L}_{P1}$ and $\rm{L}_{P2}$. The overall $\mathcal{B}$-field shift in the experiment is $\Delta/(2\pi)\sim\!\times 12$ GHz.}
	\label{fig:BFieldWithMiniShims}
\end{figure}

We measure the axial $\mathcal{B}$-field and its homogeneity using signals from the molecules. As discussed above, when a spatially narrow $\mathcal{E}$-field pulse is applied with its center at location $z_k$, the population transfer between the two levels depends on the $\mathcal{B}$-field at that location. We place $\mathcal{E}$-field pulses centered at each IR electrode ring gap $z_k, (k=1,...,31)$, and record signals as a function of applied field $\mathcal{B}_0$ in order to find the location of the level crossing, corresponding to the signal maximum. We define the associated detuning at position $z_k$, $\Delta_k$, as  ${\Delta}_k = \delta\Delta_0 + 2 {\mu}_B ({\mathcal{B}}_k -{\mathcal{B}}_0)$. Here $\delta\Delta_0$ is a small offset that depends on the quantum numbers $m_I, m^\prime_N, m^\prime_I$ of the level crossing,  ${\mu}_B$ is the Bohr magneton, ${\mathcal{B}}_k$ is the actual $\mathcal{B}$-field at $z_k$, and ${\mathcal{B}}_0$ is the applied field required for the maximum signal. We then shim the $\mathcal{B}$-field to minimize the variance within the set of $\Delta_k$ values, i.e. to make the field uniform. This shimming procedure based on molecular signals is begun only after the field has been shimmed as much as possible using signals from the NMR probe array. Then, using these level-crossing signals, we shim first with the RT shim coils, and then, in the last step, employ the mini-shim coils. For $\delta\mathcal{B}/\mathcal{B} = 10^{-7}$, the detuning $\Delta/(2\pi)$ changes by $\approx\! 1$ kHz. We routinely achieve $\delta\mathcal{B}/\mathcal{B} \lesssim 2\times 10^{-8}$ (r.m.s.) after shimming, in an overall field of ${\sim\!4600}$ Gauss (see Fig.~\ref{fig:BFieldWithMiniShims}).

\subsection{Stray Electric Field Measurement}
\label{EnrMeasurement}

It is known from previous experiments based on the same measurement concept that stray, non-reversing electric fields (${\mathcal{E}}_{nr}$), in combination with magnetic field gradients ($\partial B/\partial z$), could mimic a NSD-PV signal~\cite{Nguyen}. (By ``non-reversing'', we mean fields that do not change when we reverse the voltages used to generate the $\mathcal{E}$-field.) To minimize systematic errors resulting from such combinations, we developed a method for measuring and shimming away ${\mathcal{E}}_{nr}$. Here, we present the experimental measurement protocol and how $\mathcal{E}_{nr}$ is extracted from measured signals.
Then, we discuss various features of the extracted $\mathcal{E}_{nr}(z)$ functions and assess the accuracy of our procedure in reproducing correctly a known, applied value of $\mathcal{E}_{nr}$.

A key feature of our approach is that we assume the non-reversing $\mathcal{E}$-field can be written as a superposition of unipolar field pulses of the form given in Eqn.~\ref{UniEr}. That is, we write
\begin{equation}
\mathcal{E}_{nr}(z) = \sum_{k=1}^{31} c_k \mathcal{E}^u(z ;z_k), \label{eqn:Enrsum}
\end{equation}
so that finding $\mathcal{E}_{nr}(z)$ is equivalent to finding the set of coefficients $\{c_k\}$, where $k = 1-31$ indexes a ring gap number.  Physically, this would be an exact description of  $\mathcal{E}_{nr}(z)$ if the electrode rings were perfect equipotentials.  Moreover, any field we can apply in our system has this form, so this is a realistic operational definition of any field that is useful to define.  The set of unipolar pulses is thus used as an effective non-orthonormal (but assumed complete) basis set, from which the function $\mathcal{E}_{nr}(z)$ is composed.

\subsubsection{Basic Measurement Strategy}
\label{ExtractedEnr}
Stray non-reversing $\mathcal{E}$-fields in our apparatus are too small to cause, on their own, a measurable population transfer between the two opposite parity levels. We thus utilize a Stark interference method to amplify the effect of $\mathcal{E}_{nr}$ fields in the system to a measurable level. In addition to the unknown stray field $\mathcal{E}_{nr}$, we deliberately apply a known, larger, reversible field pulse, $\pm{\mathcal{E}}_{r}(t)$. In practice, the reversible field pulse is applied in the form of a unipolar pulse, so that $\pm{\mathcal{E}}_{r}(t) \propto \pm\mathcal{E}^u(z/v; z_k/v)$ for some gap location $z_k$ (always with peak amplitude $\mathcal{E}_{r0} = \mathcal{E}^u_0(\delta V = 0.7$ V)).  In the presence of both this reversible pulse and the non-reversing field, the total field is $\mathcal{E}(t)={\mathcal{E}}_{nr}(t) \pm {\mathcal{E}}_{r}(t)$. Following Eqn.~\ref{eqn:signalEPulse}, the resultant signal is 
\begin{eqnarray}
S_\pm(\Delta)&\propto& |\tilde{\cal{E}}_{r}(\Delta)|^2+ |\tilde{\cal{E}}_{nr}(\Delta)|^2\nonumber\\
&& \pm
\tilde{\cal{E}}_{r}^\ast(\Delta)\tilde{\cal{E}}_{nr}(\Delta)
\pm \tilde{\cal{E}}_{r}(\Delta)\tilde{\cal{E}}_{nr}^\ast(\Delta).
\end{eqnarray} 
The difference between signals taken with $\pm\mathcal{E}_r$, $S_\delta(\Delta)\equiv S_+(\Delta)-S_-(\Delta)$, arises from the interference terms: 
\begin{equation}
S_\delta(\Delta)\propto\left(\tilde{\cal{E}}_{r}^\ast(\Delta)\tilde{\cal{E}}_{nr}(\Delta)+
\tilde{\cal{E}}_{r}(\Delta)\tilde{\cal{E}}_{nr}^\ast(\Delta)\right),
\end{equation}
where the unknown parameters are the real and the imaginary parts of $\tilde{\mathcal{E}}_{nr}(\Delta)$. 

With two unknown parameters, this is an underdetermined system. Therefore, both measurements, $S_\pm(\Delta)$, are repeated with the applied reversible field ${\mathcal{E}}_r\left(t\right)$ shifted by a time offset $t_i$, i.e. ${\mathcal{E}}_r\left(t-t_i\right)$. In practice, this is done by applying the unipolar pulse at a different gap position.  By the shift theorem of Fourier transforms, $\mathcal{F}\left[\mathcal{E}\left(t-t_i\right) \right]=e^{-i\Delta t_i}\mathcal{F}\left[\mathcal{E}\left(t\right) \right]=e^{-i\Delta t_i}\tilde{\mathcal{E}}(\Delta)$, where $\mathcal{F}$ denotes the Fourier transform. Now, the signal difference is
\begin{equation}
S_\delta(\Delta, t_i)\propto e^{i\Delta t_i}\tilde{\mathcal{E}}_{r}^\ast(\Delta)\tilde{\mathcal{E}}_{nr}(\Delta)+ e^{-i\Delta t_i}\tilde{\mathcal{E}}_{r}(\Delta)\tilde{\mathcal{E}}_{nr}^\ast(\Delta).
\end{equation} 
Since the unipolar pulse $\mathcal{E}_r\left(t\right)$, when centered at $t=0$, is a real and even function, its Fourier transform $\tilde{\mathcal{E}}_r\left(\Delta\right)$ is real and even: $\tilde{\mathcal{E}}_r^\ast\left(\Delta\right)=\tilde{\mathcal{E}}_r\left(\Delta\right)$. Then, defining $\alpha(\Delta)\equiv \mathrm{Re}[\tilde{\mathcal{E}}_{nr}(\Delta)]$ and $\beta(\Delta)\equiv \mathrm{Im}[\tilde{\mathcal{E}}_{nr}(\Delta)]$, the signal difference becomes
\begin{equation}
S_\delta(\Delta, t_i)\propto  \tilde{\mathcal{E}}_{r}(\Delta)\left[ \alpha(\Delta)\cos(\Delta t_i)
+ \beta(\Delta)\sin(\Delta t_i)\right].
\label{Equ:EnrFinalDiff}
\end{equation}

To determine $\alpha(\Delta)$ and $\beta(\Delta)$ and hence extract $\mathcal{E}_{nr}(t)$, the following procedure is employed. We measure $S_\delta(\Delta, t_i)$ at a regularly spaced set of detunings $\Delta = \Delta_k = k\Delta_0$, where $k = -N, -N+1,... N-1, N$.   In principle, to determine both functions $\alpha$ and $\beta$, it suffices to obtain data across the range of detunings $\Delta = {\Delta_k}$ for only two values of the reversible field pulse center, $t_i$. However, for redundancy we use a set of $t_i$ values, denoted as $\{t_i\}$, that contains at least three different values.  This yields a set of overconstrained equations for $\alpha(\Delta_k)$ and $\beta(\Delta_k)$, for each $k$.  The best values of $\alpha(\Delta_k)$ and $\beta(\Delta_k)$ are assigned by minimizing the r.m.s. deviation to the data in this overconstrained set. The time-domain form of the unknown non-reversing field is then calculated from the Fourier series:
\begin{equation}
\mathcal{E}_{nr}(t)=\frac{\Delta_0}{\sqrt{2\pi}}\sum_{k=-N}^{N}[\alpha(k\Delta_0)+i\beta(k\Delta_0)]e^{ik\Delta_0 t}.
\label{FourierSeriesE}
\end{equation}
Like any function, $\mathcal{E}_{nr}(t)$ can be written as the sum of an even and an odd part: $\mathcal{E}_{nr}(t)=\mathcal{E}_{nr}^{even}(t)+\mathcal{E}_{nr}^{odd}(t)$. As noted before, the Fourier transform of the even part is real and even, while the Fourier transform of the odd part is imaginary and odd. Using these symmetry properties, we reduce noise in the data by replacing $\alpha$ and $\beta$ in Eqn.~\ref{FourierSeriesE} with the symmetrized values $\bar{\alpha}(\pm k\Delta_0) = [\alpha(k\Delta_0)+\alpha(-k\Delta_0)]/2$ and $\bar{\beta}(\pm k\Delta_0) = \pm[\beta(k\Delta_0)-\beta(-k\Delta_0)]/2$. Finally, we take the inverse Fourier transform of $\tilde{\mathcal{E}}_{nr}(\Delta) = \alpha(\Delta) + i\beta(\Delta)$ to find the non-reversing field $\mathcal{E}_{nr}(t)$.

This completes the logical connection between ideal data generated while applying the known reversible fields $\pm \mathcal{E}_{r}$, and the unknown, stray non-reversing field $\mathcal{E}_{nr}$ that we wish to determine. In principle, this relation should be exact, and should be independent of the set of values $\{t_i\}$ (temporal locations of the center of the reversing field pulses). 
However, by deliberately applying $\mathcal{E}_{nr}$-fields with known forms, it was quickly observed that the functions $\mathcal{E}_{nr}(z)$ determined by simple application of Eqns.~\ref{Equ:EnrFinalDiff} and \ref{FourierSeriesE} had significant qualitative differences from the input fields.  This was found to be due to a variety of effects such as spectral windowing artifacts  (due to the finite range of detunings employed in our procedure), relative dephasing between interfering amplitudes from widely-separated fields (due primarily to the finite longitudinal velocity spread of the molecular beam), distortions due to small magnetic field inhomogeneties, etc. 

To account for these effects, we developed a complex, multi-step procedure for accurately determining the ambient non-reversing fields $\mathcal{E}_{nr} (z)$.  The details of this procedure and the logic behind it are given in Appendix \ref{Effects on Enr}.  Even with our optimized procedure, small but noticeable differences were observed between the fields assigned in this way, and known forms of deliberate, non-reversing input fields.  Hence we distinguish between the ``assigned'' non-reversing field,  $\mathcal{A}_{nr}(z)$, as determined from this procedure, and the physical stray field $\mathcal{E}_{nr} (z)$.    Just as for $\mathcal{E}_{nr} (z)$, we define $\mathcal{A}_{nr}(z)$ as a field that can be generated by applying voltages to the electrodes in our system, i.e. as a superposition of unipolar field pulses.

The accuracy of match between $\mathcal{A}_{nr}(z)$ as determined via this procedure, and the actual underlying field $\mathcal{E}_{nr}(z)$, was verified experimentally.  To do this, we applied deliberate non-reversing unipolar field pulses as input to the system. The resulting output ``assigned'' field matched the input consistently well, with the magnitude of residual differences always satisfying $|\mathcal{A}_{nr}(z) - \mathcal{E}_{nr}(z)| < 15$ mV/cm.  Similarly good agreement was found by applying the extraction procedure to simulated data with more complex forms of $\mathcal{E}_{nr}(z)$, e.g., superpositions of several unipolar pulses. The agreement remained good even when the input form of the field $\mathcal{E}_{nr}(z)$ was not constrained to be a sum of unipolar pulses centered on ring gap locations $z_k$, i.e., when the simulated input field included contributions that would correspond physically to the electrode rings \textit{not} being equipotentials, as for example if there were patch potentials on the electrode surfaces.


\subsubsection{Experimental Measurement and Shimming of $\mathcal{E}_{nr}$}
\label{AnrExperimentData}

To nullify the ambient non-reversing field $\mathcal{E}_{nr}(z)$ that exists in the apparatus, we apply voltages to the IR electrodes that create a field meant to cancel the assigned non-reversing field $\mathcal{A}_{nr}(z)$.  Since $\mathcal{A}_{nr}(z)$ is, by construction, a superposition of fields that can be generated by applying voltages to the electrodes in our system, it is straightforward to apply shim voltages to each electrode to nullify each term in the superposition.

\begin{figure}
\includegraphics[width=83mm]{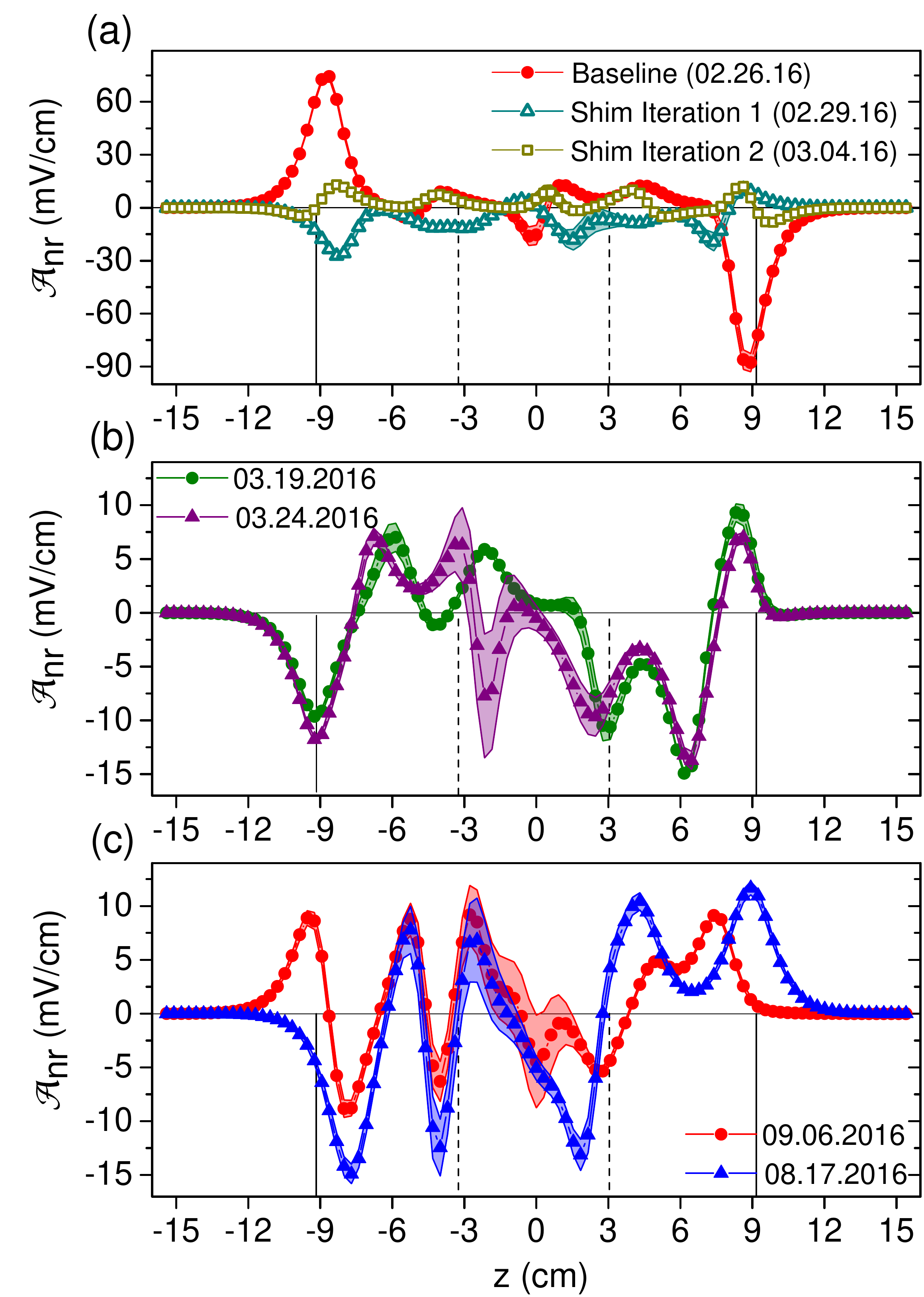}
\caption{(color online) Experimentally determined values of the ambient non-reversing field $\mathcal{A}_{nr}(z)$ and reproducibility of measurements after shimming. Points indicate best values and shaded areas indicate ranges of $\pm 1 \sigma$ statistical uncertainty. Legends indicate the date on which data was taken. (a) Suppression of $\mathcal{E}_{nr}$ by shimming. After two consecutive iterations of shimming (dark green and blue green) beginning from the baseline measurement (red), the residual assigned stray field has magnitude $|\mathcal{A}_{nr}(z)|< 15$ mV/cm everywhere within the IR. (b-c) Reproducibility of shimmed $\mathcal{A}_{nr}$ measurements taken with the same shim voltage values many days apart. Typically $\langle\mathcal{A}_{nr}(z)\rangle_{\rm r.m.s.}< 6$ mV/cm  in the region between lasers $L_{\rm P1}$ and $L_{\rm P2}$ (dashed vertical lines).  Solid lines denote the region of the IR where electrodes control the $\mathcal{E}$-field.}
\label{fig:AnrSameIterCompared}
\end{figure}

After two consecutive iterations of $\mathcal{E}_{nr}$ shimming, we find that the remaining non-reversing field consistently has magnitude $|\mathcal{A}_{nr}(z) |< 15$ mV/cm. We observed, both in numerical simulations and experimental data taken with known deliberately applied non-reversing fields, that further shim iterations did not consistently reduce the magnitude of the assigned stray field. This is believed to be due to the fact that, after two shimming iterations, the residual non-reversing field has only high frequency components that are not captured by the $\mathcal{A}_{nr}$ assignment algorithm. However, we also observed that further iterations of shimming did not \textit{increase} the magnitude of the assigned residual field.  Hence, repeated shimming after the 2$^{\rm nd}$ iteration serves to change the functional form of $\mathcal{A}_{nr}(z)$ in an effectively random manner, without increasing its typical magnitude. We take advantage of this property as a way to effectively randomize the stray non-reversing field $\mathcal{E}_{nr}(z)$, while keeping its magnitude as small as it is possible for us to determine. Fig.~\ref{fig:AnrSameIterCompared}(a) shows typical measured $\mathcal{A}_{nr}$ functions under various conditions.

Additionally, we investigated both long and short time scale drifts in the residual $\mathcal{A}_{nr}$ after shimming. Fig.~\ref{fig:AnrSameIterCompared}(b)/(c) displays measurements taken many days apart with the same shim voltage values. We observed no drifts in $\mathcal{A}_{nr}$ above the range of $\sim\!15$ mV/cm over this time scale, which was typical of the time used to collect NSD-PV data. Even though the exact shape of $\mathcal{A}_{nr}(z)$ is not reproducible from run to run over long periods of time, our measurements of the non-reversing field after at least two iterations of shimming consistently have peak magnitude $|\mathcal{A}_{nr}|<15$ mV/cm and r.m.s. variation $\langle \mathcal{A}_{nr}\rangle_{\rm r.m.s.} < 6$ mV/cm, in the entire region between the parity state projection lasers.


\section{NSD-PV MEASUREMENT}
In this section, we describe the details of NSD-PV data collection and how we deduce the value of the weak matrix element, $W$, from fluorescence data. Before each NSD-PV run, using the methods described in Sec.~\ref{SystemCharacterization}, we (1) measure and shim the $\mathcal{B}$-field, and then measure the shimmed $\mathcal{B}$-field; (2) measure and shim the non-reversing $\mathcal{E}$-field, and then measure the shimmed non-reversing $\mathcal{E}$-field. We proceed to NSD-PV measurements after verifying that the $\mathcal{B}$-field homogeneity is $(\delta\mathcal{B}/\mathcal{B})_{\rm{r.m.s.}}<0.05$ ppm, and $|\mathcal{A}_{nr}(z)|\leq 15$ mV/cm everywhere within the IR, i.e. everywhere between gap 1 and gap 31.

\subsection{NSD-PV Data Structure}
\label{NSD-PV Data Structure}

Parity violation data is grouped into three segments: pulses, blocks and runs. For a single pulse of the molecular beam, we record signals from the PMT, as well as all 32 electrode voltages (to ensure fidelity of electrical connections) and signals from the photodiodes for both parity state projection lasers (to verify shutter states, open or closed). Electrode voltages are changed on every pulse to one of the five voltage set and shutter state conditions (see Sec.~\ref{PrincipleExp}). Pulses are marked as bad and excluded from the data analysis when: (1) any of the measured electrode voltages do not match the intended applied voltages, (2) a photodiode signal does not match the expectation for the intended shutter state, (3) any of the laser frequencies deviate significantly from their set points, or (4) when the detected molecule counts are below a pre-set threshold (in which case a stepper motor turns the barium wheel in the source chamber, changing the surface exposed to the ablation laser until the signal size reaches a pre-set, typical size). Most bad pulses resulted from factor number (3); this happens typically for 10-30 out of every $300$ pulses.

A collection of pulses with the same $\mathcal{B}$-field setting (i.e., same $\Delta$) constitutes a block. Then, a block of data gives a single asymmetry data point, $\mathcal{A}(\Delta)$, and its associated uncertainty $\delta\mathcal{A}(\Delta)$. To account for cases when pulses are excluded due to errors, data in a block is acquired until a minimum of $300$ good pulses is recorded. The parity state projection laser shutters simultaneously switch between open and closed every 60 pulses. During each shutter state, all four $\mathcal{E}$-field conditions are applied, in a random order. This way we remove any correlations that might arise from $\mathcal{B}$-field drifts as we sweep over a range of $\Delta$ values. During a block, we measure the magnetic field using one of the off-axis NMR probes repeatedly, 20-30 times. At the end of a block, these values are averaged and used to assign the $\Delta$ value for the block. An entire block of data is excluded from the data analysis when any of several experimental errors (e.g. consistently low signal size, too high OPR value for either of the parity state projection lasers, etc.) is detected. In a typical run, with $\approx\!500$ blocks on average, $\sim\!30$ are assigned as bad.

A run comprises $250-500$ blocks with the same four $\mathcal{E}$-field voltage sets and common $\mathcal{B}$-field shim settings, but with an array of different values of $\Delta$. We make measurements in the detuning range $1~\text{kHz}\leq |\Delta/(2\pi)| \leq 4~\text{kHz}$, in steps of $\delta\Delta/(2\pi)\approx\!350~\text{Hz}$. Due to small fluctuations in the ambient magnetic field, we observe differences in $\Delta$ even when we try to set the same value of $\Delta$ with the Z0 RT shim coil. These detuning fluctuations are typically not larger than ${2\pi\!\times\!100}$ Hz. To account for this, we bin adjacent asymmetry points together into detuning intervals of size $\Delta_{\text{bin}}/({2\pi})=250$ Hz and calculate the weighted average and the standard error of the combined asymmetry data points within each bin. For a typical data run with $\approx\! 500$ blocks, a given bin has $N\approx\!13$ asymmetry data points. We observed no effects on the extracted value of $W$ by using different binning interval sizes.


\subsection{$W$ Measurement with $^{138}\mathrm{Ba^{19}F}$ and Statistical Uncertainty}
\label{finalNSDPVData}

\subsubsection{Asymmetry Fit Function}
\label{FitFunc}

For a PV run, the value for the NSD-PV matrix element $W$ is extracted by fitting the asymmetry data to the function $\mathcal{A}_{\rm{fit}}(\Delta)$, where
\begin{equation}
\mathcal{A}_{\rm{fit}}(\Delta)= W_{\rm{fit}}\mathcal{A}_0(\Delta) +a_0+a_1\Delta.
\label{Equ:DataFinalAsymFit}
\end{equation}
Here, $\mathcal{A}_0(\Delta)=\mathcal{A}_{thy}(\Delta) /W$, with $\mathcal{A}_{thy}(\Delta)$ defined in Eqn.~\ref{AsymParity}. The free parameters in the fit are $W_{\rm{fit}}$ as well as the auxiliary coefficients $a_0$ and $a_1$. The rationale for inclusion of these auxiliary parameters is as follows.  

First, experimental and simulated data were observed to sometimes exhibit a constant offset (independent of $\Delta$) in the asymmetry. To account for this observation, we empirically modify our fit function with the addition of a constant offset term, $a_0$. We know, from numerical simulations and analytical calculations, that such a term can be induced by $\mathcal{E}_{nr}$ fields alone (with its value dependent on the specific shape and amplitude of $\mathcal{E}_{nr}$). In these simulations and calculations, we find that the presence of a nonzero value of $a_0$ has no clear correlation with systematic errors in $W$.

The linear term in Eqn.~\ref{Equ:DataFinalAsymFit}, associated with the coefficient $a_1$, was included to account for observations in numerical simulations where various deliberate imperfections were added to the system. In particular, we observed in these simulations that a linear term typically arises from $\mathcal{E}_{nr}$ fields spatially coincident with $\mathcal{B}$-field gradients.  Moreover, in these simulations we found that a nonzero $a_1$ term \textit{is} strongly correlated with potential systematic errors in $W$, i.e. with a contribution to $\mathcal{A}_0(\Delta)$ with similar dependence on $\Delta$ as given by Eqn.~\ref{AsymParity}. Therefore, we considered it important to include this linear term in our data analysis. We treat any nonzero $a_1$ values resulting from fits to our NSD-PV data as a preliminary manifestation of a systematic error in $W$. However, inclusion of this term in all asymmetry fits comes at a cost in statistical power. This is because, over the limited range of $\Delta$ values used in our NSD-PV data sets, there is substantial covariance between fitted values of $W$ and of $a_1$. Because of this covariance, including $a_1$ as a free parameter increases the statistical uncertainty in $W$ from the fits by a factor of $\sim\!\sqrt{2}$.


\subsubsection{NSD-PV Measurements}
\label{NSDPVdata}

\begin{figure}
	\includegraphics[width=83mm]{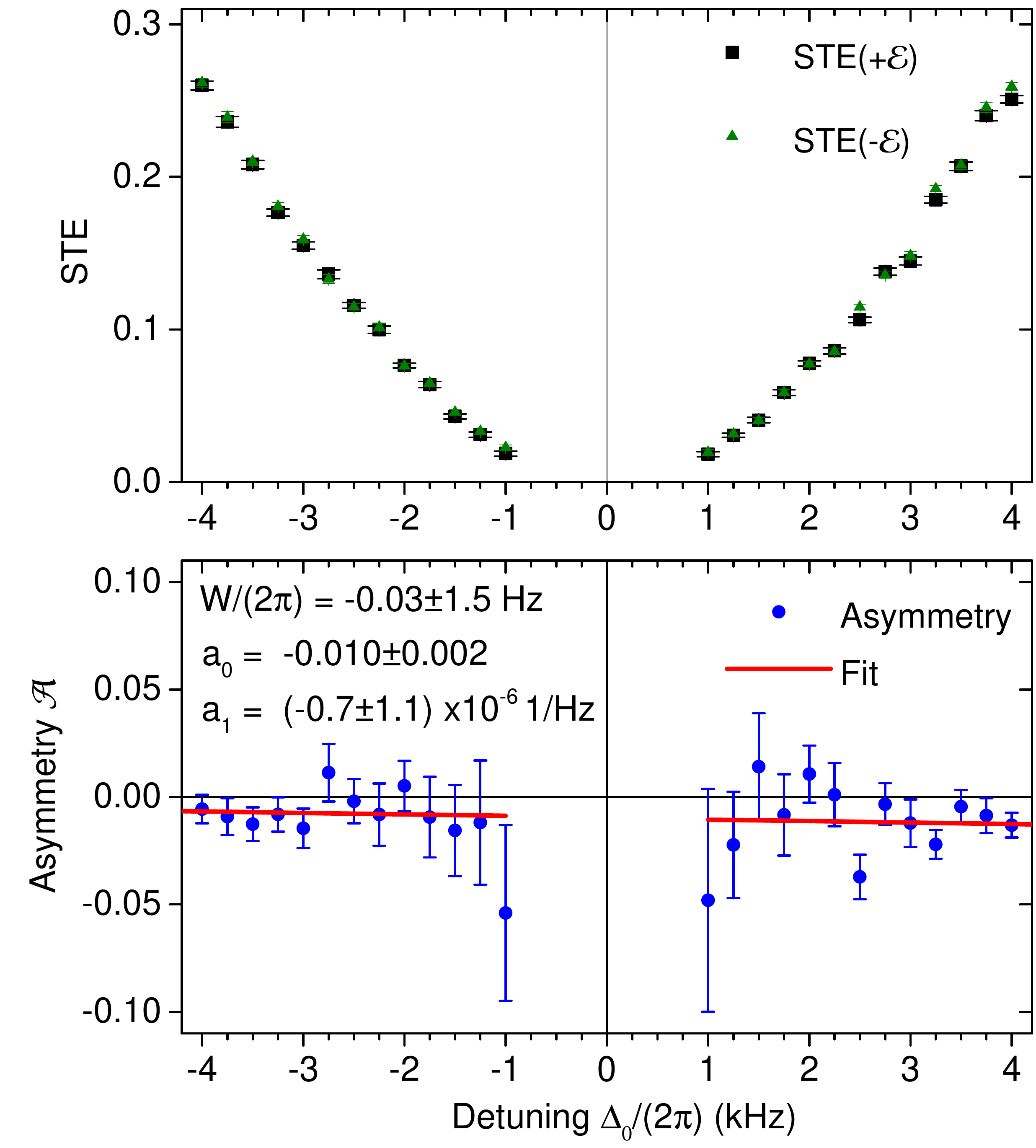}
	\caption{(color online) Sample PV data with $^{138}\mathrm{BaF}$ at crossing F (see Fig.~\ref{fig:Crossings138}). (a) Rectangles (triangles) indicate state transfer efficiency (STE) data taken with the positive (negative) value of $\mathcal{E}_0$ for the $\mathcal{E}$-field sine pulse. (b) Measured asymmetry (circles) and the fit (line) using Eqn.~\ref{Equ:DataFinalAsymFit}. For this run, we acquired $\approx\! 4.4$ hours of good data. The average number of molecules per pulse was $\bar{N}_0\approx\!96$, giving a total of ${N}_0\approx\! 3.0 \times 10^6$ molecules for each one of the five signal types measured (see Table~\ref{TableSignals}). Best-fit values for the parameters $W, a_0$, and $a_1$, along with their $1\sigma$ statistical uncertainties, are shown in the figure. The reduced $\chi^2$ for this fit is $\chi_\nu^2=0.94$.}
	\label{fig:AsymmetryRun3842}	
\end{figure}

\begin{figure*}	
\includegraphics[width=180mm]{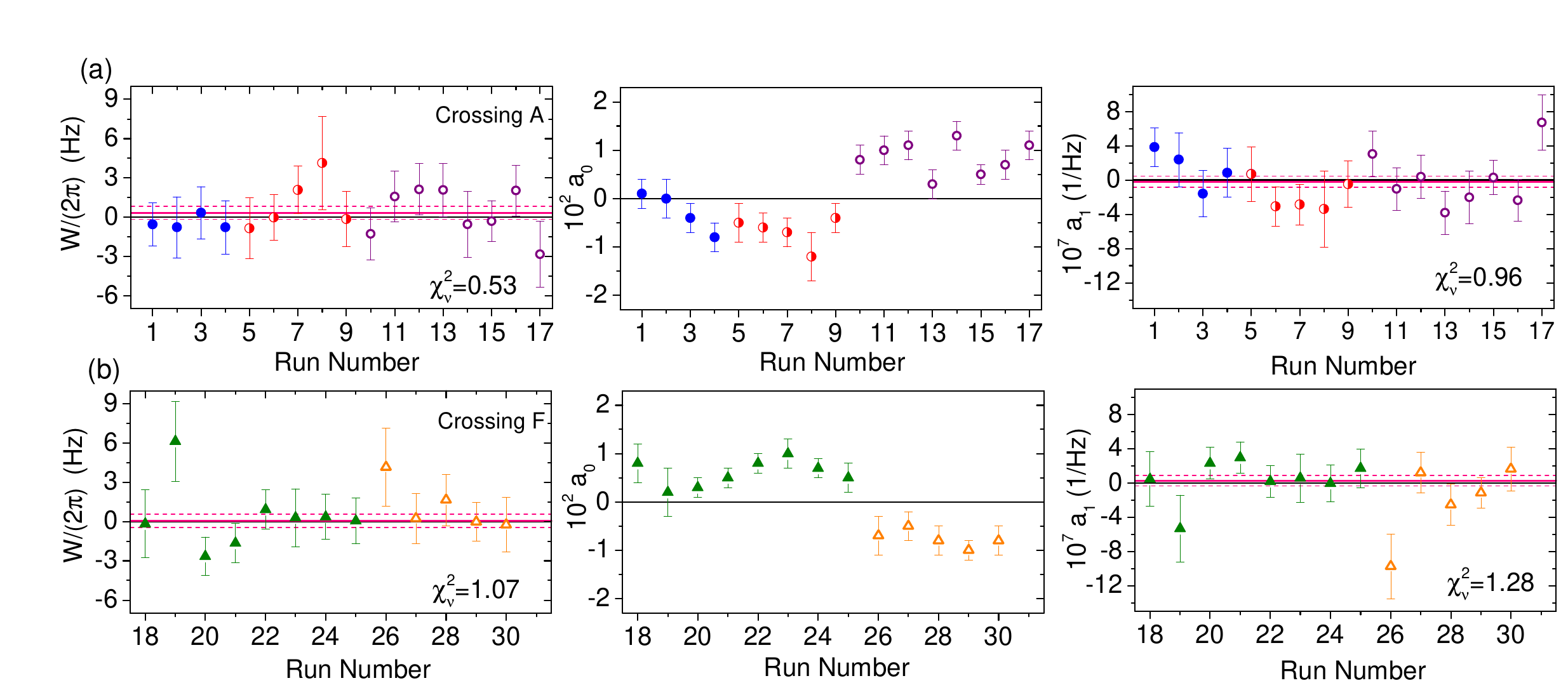}
\caption{(color online) Summary of NSD-PV data with $^{138}\mathrm{BaF}$. Data runs plotted with the same symbol were taken with the same $\mathcal{E}_{nr}$ shim voltage set. Error bars represent 1$\sigma$ statistical uncertainties. (a) and (b) show results from NSD-PV data runs at crossing A and F, respectively. Central values of weighted averages are denoted by a solid horizontal line, and the $1\sigma$ uncertainty range by dashed lines. Left: The weighted average of A crossing runs is $W^{(A)}/(2\pi) =0.32\pm0.49$ Hz, F crossing runs is $W^{(F)}/(2\pi) =0.05\pm0.51$ Hz. Middle: Asymmetry offset term, $a_0$, for each run. With each different $\mathcal{E}_{nr}$ shim voltage set, the $a_0$ value changes, as expected, while for runs with the same $\mathcal{E}_{nr}$ shim voltage set, results for $a_0$ are consistent. Right: Linear asymmetry term, $a_1$, for each run. The weighted average for A crossing runs is $a_{1}^{(A)}=(-2\pm6)\times10^{-8}$ 1/Hz, and for F crossing runs $a_{1}^{(F)}=(3\pm6)\times10^{-8}$ 1/Hz.} 
\label{fig:AllPVonlyA0BothCrossings}
\end{figure*}

Fig.~\ref{fig:AsymmetryRun3842} shows STE signals and asymmetry from a typical run with $^{138}\mathrm{BaF}$. We fit asymmetry data to Eqn.~\ref{Equ:DataFinalAsymFit} with parameter values $T_e=2\pi/\omega=87~\mu$s, $T_{f1}=7.4~\mu$s, $T_{f2}=8.9~\mu$s, $\mathcal{E}_0=1$ V/cm, $\omega/(2\pi)=11.4$ kHz and $d/(2\pi)=3360~(3530)$ Hz/(V/cm) for the A (F) crossing (see Figure~\ref{fig:Crossings138}). Fig.~\ref{fig:AllPVonlyA0BothCrossings} shows the results for $W$, $a_0$, and $a_1$ for all parity violation data with $^{138}\mathrm{BaF}$. We performed NSD-PV measurements with several different $\mathcal{E}$-field shim voltage values, in order to see any dependence on a specific form of the ambient $\mathcal{E}_{nr}$ field. Here, by ``different" we mean that the $\mathcal{A}_{nr}$ signals have different, effectively random shapes but are all consistent with zero within our measurement accuracy, i.e., $|\mathcal{A}_{nr}(z)|<\!15$ mV/cm everywhere within the IR. We generate this effectively random difference in the ambient $\mathcal{E}_{nr}$ after two consecutive shim iterations by shimming for a third time (or a fourth time and so on). As seen in Fig.~\ref{fig:AllPVonlyA0BothCrossings}(a), measured $W$ values for different $\mathcal{E}_{nr}$ fields (shown in different colors) are consistent, as expected. Data runs taken with the same $\mathcal{E}_{nr}$ field shim set yield consistent $a_0$ values (within $2\sigma$), but, as expected (since $a_0$ depends on the stray non-reversing field), the values of $a_0$ differ for different shim voltage values. The fitted $a_1$ values are consistent with $0$, consistent with the hypothesis that there are no significant systematic errors in $W$.

As a further test for systematics, we made measurements at both level crossings, A and F, as shown in Figure~\ref{fig:Crossings138}. The angular momentum dependence of  $H^{\rm eff}_{\rm P}$, via the operator $C\equiv\left(\bm{n} \times \bm{S}\right)\cdot \bm{I}/{I}$, ensures that the matrix element $W$ has a different value at each crossing (see Eqn.~\ref{Eqn.W2}). Note that $C$ is a purely angular factor, whose matrix elements $\tilde{C}$ depend on the angular momentum quantum numbers of both states at a given crossing. The values of these dimensionless matrix elements in our case are $\tilde{C}_A=-0.41i$ and $\tilde{C}_F=0.39i$, at crossings A and F respectively. By contrast, the dipole matrix element $d$ for each crossing has a similar value and the same sign: $d=d_A=3360$ Hz/(V/cm) for crossing A and $d=d_F=3530$ Hz/(V/cm) for crossing F. This means that systematics due to a common $\mathcal{E}_{nr}$ field would give asymmetries of the same sign at both crossings, while a true NSD-PV signal would cause the asymmetry to reverse sign but have nearly identical magnitude. As a result, nearby level crossings within the same molecule, such as the A/F pair in $^{138}$BaF, are powerful tests for systematic errors. NSD-PV measurements at the A and F crossings are shown in Fig.~\ref{fig:AllPVonlyA0BothCrossings}(a) and (b), respectively. Our final results for W at each crossing are $W^{(A)}/(2\pi) =0.32\pm0.49$ Hz and $W^{(F)}/(2\pi) =0.05\pm0.51$ Hz, where the uncertainties here are only statistical.

\section{SYSTEMATIC ERRORS}
Systematic errors, were a principal concern in the design of our experiment and development of the NSD-PV measurement scheme. In previous experiments  using a similar scheme but with Dy atoms~\cite{Nguyen}, it was observed that stray, non-reversing electric fields ($\mathcal{E}_{nr}$) in combination with magnetic field gradients ($\partial \mathcal{B}/\partial z$) could mimic the NSD-PV signal. To minimize the systematic contributions resulting from such combinations, we established the methods detailed in Sec.~\ref{SystemCharacterization} for measuring and then shimming away to zero, as accurately as possible, both $\partial\mathcal{B} / \partial z$ and $\mathcal{E}_{nr}$. 

This section discusses how the magnitude and uncertainty of possible systematic errors in the NSD-PV weak matrix element, $W$, are determined. First, we outline our strategy for identifying parameters that produce systematic errors. Then, we present the results of systematic error search measurements conducted with deliberately amplified  $\mathcal{E}_{nr}$ fields, $\mathcal{B}$-field inhomogeneities, and/or
offsets in laser detunings. We conclude with our assignment of the final systematic error shift, and its associated uncertainty, in the measurement of $W$ for $^{138}$BaF.

\subsection{Strategy for Determining Systematic Errors}
\label{SysOverallValue}

Our strategy for identifying systematic errors is as follows. We first shim imperfections as described above, and set upper bounds on their residual nonzero values. Then, we intentionally amplify an experimental imperfection via an associated control parameter, and make a NSD-PV measurement. We analyze the resultant asymmetry signal as we do for a regular NSD-PV run and define the result for $W$ as $W^{+d}$, where the superscript $+d$ signifies use of an imperfection-controlling parameter with a positive sign. Next, we reverse the sign of the imperfection-controlling parameter, then make another NSD-PV measurement to obtain $W^{-d}$.

The results of these measurements are evaluated using the following criteria. If ${(W^{+d} - W^{-d})/2}$ is within $3\sigma$ of the null result, we conclude that there is no definitive systematic contribution associated with the parameter being varied. If, on the other hand, $(W^{+d} - W^{-d})/2 >3\sigma$, we treat the result as suspicious and make additional measurements while changing the same parameter. If the results from these additional measurements also yield $(W^{+d} - W^{-d})/2>3\sigma$, we conclude that a mechanism generating a clear systematic offset has been identified. 

In our final systematic error budget, we include contributions to the uncertainty due to the systematic error associated with an imperfection-control parameter if either one of the following criteria are met: (1) the parameter has been observed to cause a systematic shift in $W$ in a previous NSD-PV experiment employing similar measurement techniques to ours; or (2) the parameter produced a definitive change in $W$, as described above. However, we include a systematic shift in our measured $W$ value --- that is, a true systematic error, rather than an uncertainty in a possible systematic error --- only when criterion (2) is met.

\subsection{Deliberate $\mathcal{E}_{nr}$ Fields}
\label{SysEnr} 

\begin{figure}
\includegraphics[scale=0.355]{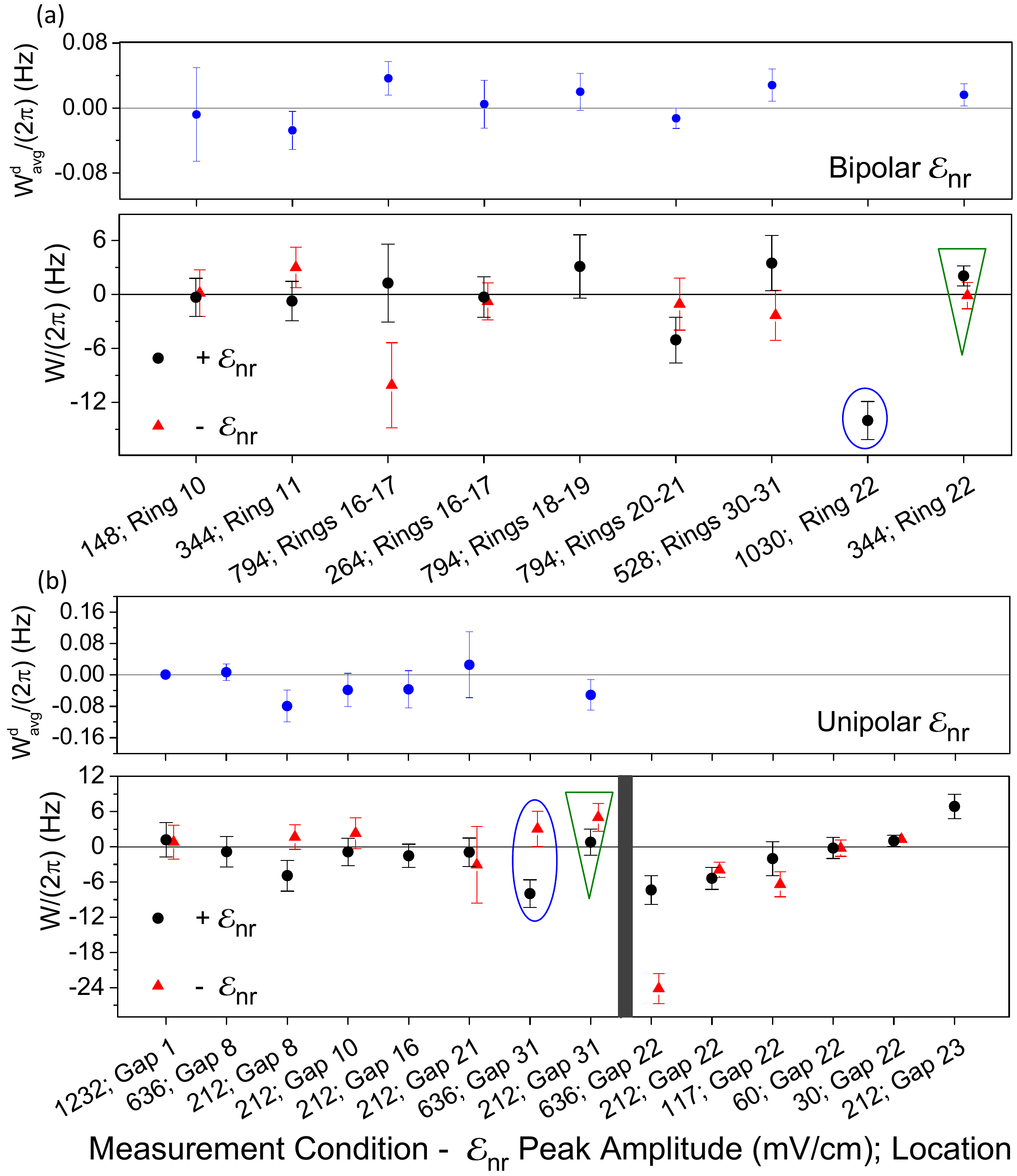}
\caption{(color online) Systematic error search measurements made with intentionally amplified $\mathcal{E}_{nr}$ fields. (a)/(b) Measurements made with deliberately exaggerated bipolar/unipolar field pulses. $x$-axes specify the center location and peak amplitude $\mathcal{E}_{0}^{b/u}$ (in mV/cm) of the applied field pulse. Top: The value $W^d_{\rm avg} = (W^{+d}-W^{-d})/(2A_d)$. Bottom: Results for $W^{+d}$ and $W^{-d}$, i.e., measurements made with $\mathcal{E}_{nr}$ pulses with the same location and magnitude, but amplitude reversed in sign. For reference, the typical ambient $\mathcal{E}_{nr}$ r.m.s. amplitude is ${\mathcal{E}_{nr0}^{a}=5.1~\rm mV/cm}$ under normal NSD-PV run conditions. Data points indicated with blue ellipsoids did not reproduce when repeated (green triangles) with somewhat smaller values of the deliberate $\mathcal{E}_{nr}$ field pulse, and so are not included in the final systematic error budget. In (b), results shown to the right of the vertical gray band are from measurements made with a deliberate unipolar pulse centered near the location of the $L_{\rm P2}$ laser beam. These are treated separately (see Sec.~\ref{SysLaserDetuning}) since they were found to be related to a different mechanism.}
\label{fig:DeliberateEnr}
\end{figure}

We use the IR electrodes to generate deliberate $\mathcal{E}_{nr}(z)$ field shapes that could potentially lead to systematic errors. We employ two different shapes of $\mathcal{E}_{nr}$ fields in these systematic error search measurements: the unipolar pulse, defined in Eqn.~\ref{UniEr}, and the bipolar pulse defined in Eqn.~\ref{BiEr}. We divide the IR into 5 sections based on the locations of different steps employed in the NSD-PV measurement sequence: before/after the 1$^{\rm st}$/2$^{\rm nd}$ prism rings, the prism rings themselves, and the central region where the sinusoidal $\mathcal{E}$-field is applied. We chose a gap location (a ring) in each section and intentionally exaggerated the $\mathcal{E}_{nr}$ at that location by applying a voltage step at that gap (a voltage at that ring) to generate a unipolar (bipolar) $\mathcal{E}$-field pulse considerably larger than the size of the ambient $\mathcal{E}_{nr}$ at that location. Fig.~\ref{fig:DeliberateEnr} shows the center location and peak amplitude of the applied field pulse for each such measurement, and the corresponding $W^{\pm d}$ values. Excluding the measurements made with unipolar pulses centered at gaps 22 and 23 (which are discussed later, separately, in Sec.~\ref{SysLaserDetuning}), all of the $W^{\pm d}$ values are within $2.1\sigma$ of $0$, i.e. they cause no definitive change in the $W$ value. Furthermore, in none of these measurements did we observe a significant correlation between $W^d$ and the sign of the intentionally amplified $\mathcal{E}_{nr}$, as would be expected for a true systematic offset in $W$.

\subsection{Deliberate $\mathcal{B}$-Field Inhomogeneities}
\label{SysBField}

The RT shim coils of our primary magnet, along with the ``mini-shim" coils, allow for a variety of magnetic field gradients to be applied. We used this control to search for systematic error contributions resulting from the couplings between deliberately amplified magnetic field inhomogeneities and any ambient imperfections in the experiment. 

We investigated possible systematic errors with 3 different room temperature shim coils. These coils, referred to as Z1, Z2, and Z3, give magnetic field contributions described by the functions $\mathcal{B}_{Z1}(z)=A^{Z1}_1(z-z_{01})+A^{Z1}_0$, $
\mathcal{B}_{Z2}(z)=A_2^{Z2}(z-z_{02})^2+A_0^{Z2}$, etc. Here,the coefficients $A_k^{\rm Zn}$ and $z_{0n}$ are values determined using data from molecular level-crossing signals (as in Sec.~\ref{BfieldMeasure}), when a known current is applied to each of these coils. Then, applying a known current to one of these coils yields a known, deliberately exaggerated $\mathcal{B}$-field imperfection. We made similar measurements using various mini-shim coils.

\begin{figure*}
\includegraphics[width=180mm]{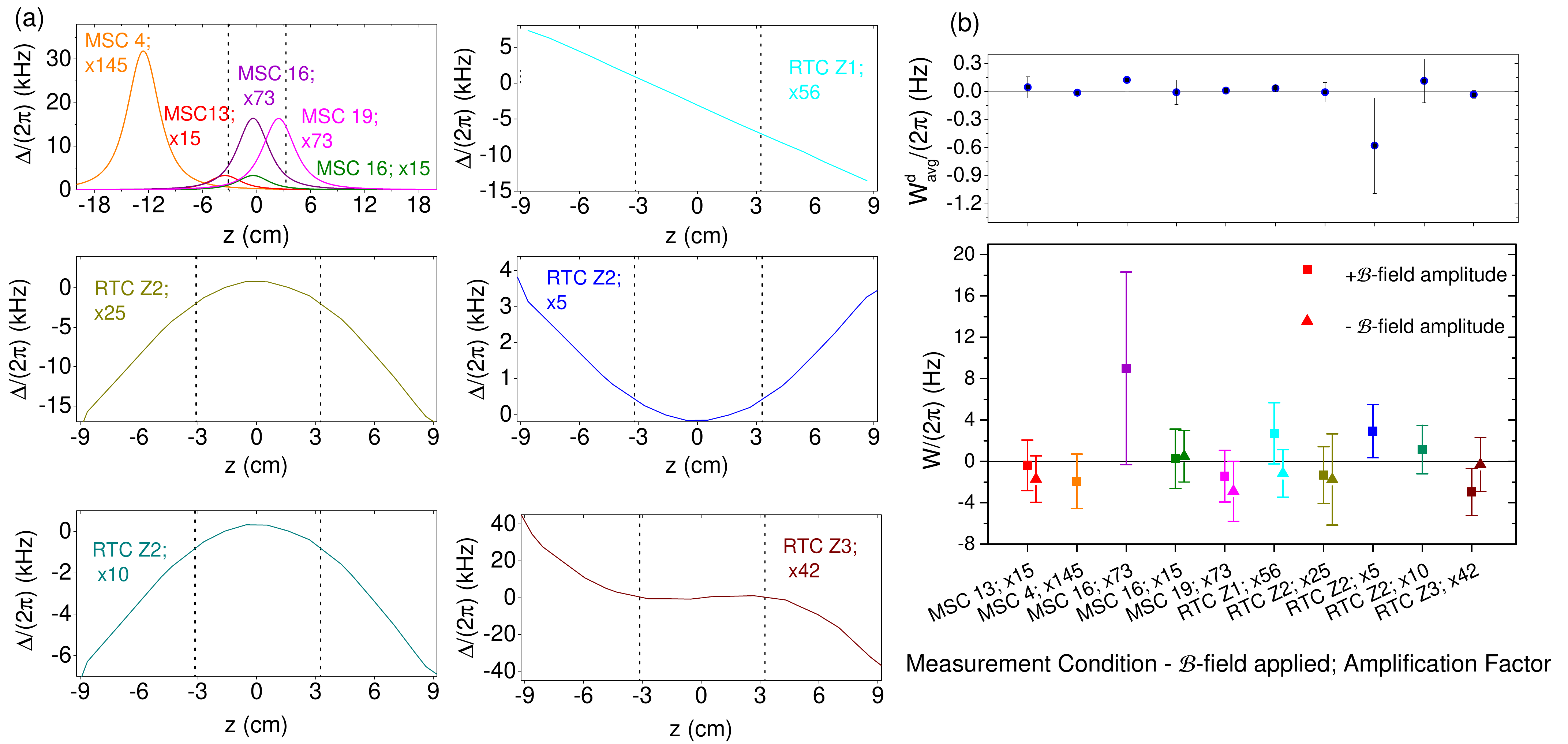}
\caption{(color online) Systematic error search measurements with intentionally amplified $\mathcal{B}$-field inhomogeneities. (a) Deliberately exaggerated $\mathcal{B}$-fields employed in the measurements summarized in (b). $\mathcal{B}$-field conditions are color-coded with the results in (b). Amplification factors are calculated using the typical standard deviation in the ambient $\mathcal{B}$-field, $\delta \mathcal{B}^a\approx\!0.07$ mGauss, i.e. $\delta\mathcal{B}^{a}*2\mu_B/(2\pi)=\delta \Delta^a/(2\pi)=0.21$ kHz, during NSD-PV runs. Dashed vertical lines represent the parity state projection laser beam locations. (b) Top: The value $W^d_{\rm avg} = (W^{+d}-W^{-d})/(2A_d)$.
Bottom: Extracted values of $W^{\pm d}$ for different amplified $\mathcal{B}$-field gradient conditions. Triangles (squares) indicate measurements made with positive (negative) applied $\mathcal{B}$-field inhomogeneity relative to the shapes shown in (a). $x$-axis labels indicate the mini-shim coil (MSC) location (for localized applied $\mathcal{B}$-field pulses) or shape (for $\mathcal{B}$-fields applied via the RT shim coils [RTCs]) of the inhomogeneity and the amplification factor $A_d$ relative to the maximum size of ambient inhomogeneities of the same shape during NSD-PV data runs. }
\label{fig:DeliberateBField}
\end{figure*}

Fig.~\ref{fig:DeliberateBField} presents the shape of the deliberately exaggerated $\mathcal{B}$-field inhomogeneity for each of these measurements, and the results obtained for the NSD-PV matrix element, $W^{\pm d}$, with each of these field imperfections applied. All measurements are within $1.3\sigma$ of the null result, and there is no significant correlation of results with the sign of the applied imperfections are the same within their uncertainties.

\subsection{Deliberate $\mathcal{E}_{nr}$ Pulse and Linear $\mathcal{B}$-field Gradient}
\label{SysAnalyticalModel}

We know from numerical simulations that certain combinations of non-reversing $\mathcal{E}$-fields and $\mathcal{B}$-field inhomogeneities mimic the NSD-PV effect and give rise to a non-zero offset in the deduced value of the weak matrix element, $W$, from our data. Here we describe a simple case of such an effect, which is amenable to analytic solution. We take the non-reversing $\mathcal{E}$-field to have a Gaussian shape, and the $\mathcal{B}$-field to have a linear gradient. We use this case for two purposes. First, by applying these field imperfections deliberately with known, exaggerated size, we demonstrate that our experiment can detect a signal with the same form as the NSD-PV signal, and replicate the expected size of the signal. Second, by using our measured limits on the size of ambient imperfections of the same form in our apparatus, we set a limit on the size of systematic errors from this type of imperfection in our actual NSD-PV data.

We define the non-reversing field as a Gaussian function and, as before, apply it in the presence of a reversible field in the form of a single cycle sine wave. That is,
\begin{align}
\mathcal{E}_{nr}(t)&=\mathcal{E}_1{e^{-\left(\frac{t}{\sigma_g}\right)}}^2,\label{EnrSys}
\end{align}
where $\mathcal{E}_1$ is the non-reversing field amplitude and $\sigma_g$ describes the width of the non-reversing $\mathcal{E}$-field. The $\mathcal{B}$-field inhomogeneity is defined such that the detuning $\Delta(t)$ has the form $\Delta(t)=\Delta_0+ \gamma*t$, where $\Delta_0$ is the uniform detuning applied in the NSD-PV measurement sequence and $ \gamma/(2\mu_B) = \partial{\mathcal{B}}/\partial t$ is the linear magnetic field gradient.

Given these definitions, we calculate the amplitude of population transfer between the nearly-degnerate states, similar to the calculation in Sec.~\ref{PrincipleExp}. Assuming $\Delta_0 , d\mathcal{E}_0 \ll \omega$, $\gamma \ll \omega^2$, and $\gamma \ll \sigma_g^{-2}$, and in the limit that the amplitude $c_+ \ll 1$ so that $1^{st}$ order perturbation theory holds, we find that the resultant ``Gaussian-sine-linear (gsl)'' asymmetry, $\mathcal{A}_{gsl}$, that mimics NSD-PV effects is given by
\begin{equation}
\mathcal{A}_{gsl}(\Delta_0)=\frac{2}{\sqrt{\pi}}\frac{\mathcal{E}_1}{\mathcal{E}_0}
\gamma\frac{\sigma_g}{2}
\left(\frac{\pi^2}{2}-3-\frac{\sigma_g^2\omega^2}{4} \right) \frac{\Delta_0}{\Delta_0^2+x^2}.
\label{AsymetryAnalyticGaussianGredient1}
\end{equation}
Here, $x=\frac{\mathcal{E}_{1}}{\mathcal{E}_{0}}\frac{\sigma_g}{2}\frac{\omega_g^2}{\sqrt{\pi}}$ is a frequency scale set by the $\mathcal{E}_{nr}$ field condition; $x \rightarrow 0$ in the limit of a vanishing non-reversing field, $\mathcal{E}_1 \rightarrow 0$. 

\subsubsection{Verification and Calibration of Sensitivity}
\label{GSLmeasurements}

\begin{figure*}
\includegraphics[width=180mm]{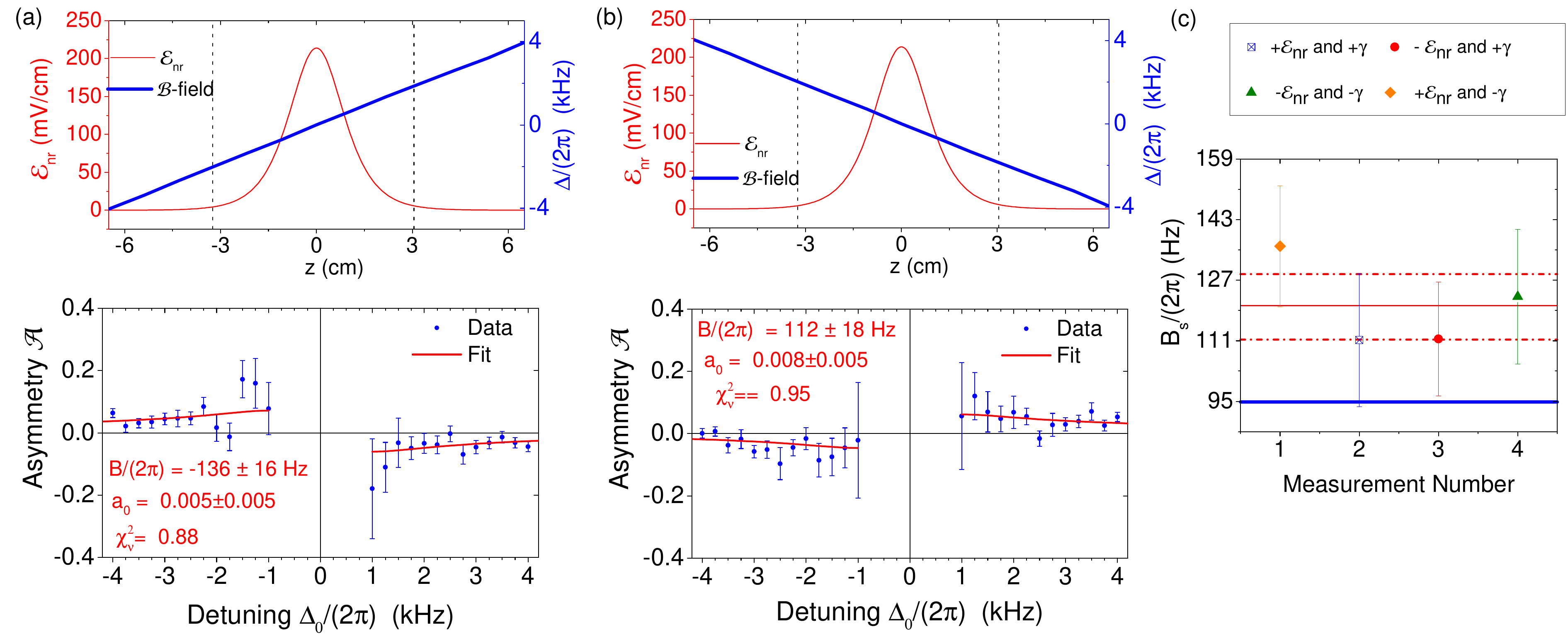}
\caption{(color online) Measurements made with simultaneously applied deliberate $\mathcal{E}_{nr}$ and linear $\mathcal{B}$-field gradient. (a-b) Top: The deliberately applied unipolar $\mathcal{E}_{nr}$ pulse centered at Gap 16 with amplification factor $A_d\approx\!42$ and the intentionally applied linear $\mathcal{B}$-field gradient with amplification factor $A_d \approx 16$. Bottom: Asymmetry data (circles) and fit (line) using Eqn.~\ref{CalibrationFitFunction} as the model function. (c) Values of the sign-adjusted asymmetry amplitude,  ${B}_s = B*sgn(\gamma)*sgn(\mathcal{E}_{nr})$, for each measurement. Horizontal solid line is the weighted average of all measurements and dot dash lines indicate the ranges of $\pm1\sigma$ statistical uncertainty. The thick solid line is the predicted value $B^c_{gsl}$.  }
\label{fig:Runs3102and3103}
\end{figure*}

To compare this simple analytical model to experimental results, we made NSD-PV measurements while applying a deliberate near-Gaussian $\mathcal{E}_{nr}$ pulse and a $\mathcal{B}$-field linear gradient. Figure~\ref{fig:Runs3102and3103} shows two such measurements performed with different signs of the $\mathcal{B}$-field gradient. We also made a second set of measurements in the same manner, but with the opposite sign of $\mathcal{E}_1$.

The resulting asymmetry data is fit to a modified version of the analytical model given in Eqn.~\ref{AsymetryAnalyticGaussianGredient1}. In particular, we fit to the function
\begin{equation}
\mathcal{A}(\Delta_0)=\frac{B*\Delta_0}{\Delta_0^2+x^2}+a_0,
\label{CalibrationFitFunction}
\end{equation}
where the fit parameters are the asymmetry amplitude, $B$, and asymmetry offset, $a_0$. Comparing to Eqn.~\ref{AsymetryAnalyticGaussianGredient1}, the predicted value for $B$ is
\begin{equation}
B_{\rm gsl}=\frac{2}{\sqrt{\pi}}\frac{\mathcal{E}_{1}}{\mathcal{E}_{0}}{\gamma\frac{\sigma_g}{2}\left( \frac{\pi^2}{2} -3 - \frac{\sigma_g^2 \omega^2}{4}\right)}.
\label{CalibrationBValue}
\end{equation}

We additionally studied numerical simulations, using the same input conditions for $\mathcal{B}(z)$, $\mathcal{E}_{nr}(z)$ and $\mathcal{E}_{r}(z)$. In these simulations, we observed that as the amplitude for the reversing $\mathcal{E}$-field, $\mathcal{E}_{0}$, and/or the non-reversing field, $\mathcal{E}_1$, increases, the agreement between the predicted value for $B$ by the analytical model, $B_{\rm gsl}$, and the one extracted from the simulated data, $B_{\text{sim}}$, deteriorated. This is due to the inadequacy of the 1$^{\rm st}$-order perturbation approximation for sufficiently large values of $\mathcal{E}_0$ and/or $\mathcal{E}_1$. In the conditions of our experimental data, the numerical simulations predicted an asymmetry $\approx\! 26\%$ smaller than expected from the analytic expression of Eqn.~\ref{CalibrationBValue}. Therefore, we modify our expectation for the value of $B$ extracted from measurements, to the corrected prediction $B_{\rm gsl}^c=C_F*B_{\rm gsl}$, where $C_F=0.74$ is the required to match the numerically simulated data for the actual values of $\mathcal{E}_{1}/\mathcal{E}_{0}$ and $\mathcal{E}_0$ used in our measurement. 

In the experimental measurements, we had $\mathcal{E}_{1}=0.212$ V/cm, $\gamma/(2\pi)=37.6$ MHz/s, $\mathcal{E}_{0}=1$ V/cm, $\sigma_g=(1.3~\rm cm)/(616~\rm {m/s}) = 21~ \mu\rm s$, and $\omega/(2\pi)= 11.4$ kHz. Then, the expected value for the fit parameter $B$ is $B_{\rm gsl}^c/(2\pi) \cong  (0.74 * 129)~\text{Hz} \cong 95$ Hz. The weighted average of measurements yields the experimental value $B_{\text{exp}}/(2\pi)= 121\pm8$ Hz. The ratio of this experimental value to the predicted value is $B_{\text{exp}}/B_{\rm gsl}^c = 1.26 \pm 0.08$, indicating that our sensitivity is calibrated correctly to within ${\sim\!25}\%$. Similar measurements, using a different functional form for the reversing $\mathcal{E}$-field and smaller magnitudes of both fields, yielded consistent results. This accuracy is sufficient for our current null measurement result with $^{138}$BaF, but in future work we will attempt to make a more accurate calibration.

\subsubsection{Associated Systematic Error in NSD-PV Data}
\label{GSLsystematics}
By comparing Eqn.~\ref{asymmetry} and Eqn.~\ref{AsymetryAnalyticGaussianGredient1}, we would deduce a nonzero value for $W$ in the presence of a combined Gaussian $\mathcal{E}_{nr}$ and linear $\mathcal{B}$-field gradient. Under our nominal NSD-PV data taking conditions, $\mathcal{E}_{1}<<\mathcal{E}_{0}$. Then, in Eqn.~\ref{AsymetryAnalyticGaussianGredient1}, $x^2\rightarrow0$, and the resulting systematic error contribution to $W$, $\Delta W^{\rm gsl}_{\rm sys}$, is given by  
\begin{equation}
\Delta W^{\rm gsl}_{\rm sys} = W^{\rm gsl}( x^2\rightarrow 0)=\frac{d\mathcal{E}_1}{2\sqrt{\pi}\omega} \gamma\sigma_g  \left[{ \frac{\pi^2}{2} -3 - \frac{\sigma_g^2 \omega^2}{4}}\right].
\label{WoffsetGSL}
\end{equation}
The typical value of ambient non-reversing $\mathcal{E}$-field localized at $z=z_{16}$ (gap 16, the center of the IR) during NSD-PV data is small and consistent with zero: $\mathcal{E}_1(\rm ambient)=\mathcal{E}_{nr}^{a}(z_{16})=- 1.5 \pm 3.5~\text{mV/cm}$, and for a localized field ${\sigma_g\approx\!21}~\mu$s, as before. The linear magnetic field gradient under normal operating conditions is also small and consistent with zero: $\gamma({\rm ambient})/v = \partial{\Delta^{a}/(2\pi)}/\partial z= 0.03\pm0.02~\text{kHz/cm}$. With these values, the systematic error shift in $W$ and its associated uncertainty, arising from this combination of imperfections, is $\Delta
W_{\rm sys}^{\rm gsl}/(2\pi)=0.01\pm 0.02 ~\text{Hz}$.


\subsection{Laser Imperfections}
\label{SysLaserDetuning}

A number of laser imperfections are possible in our experiment, such as a laser detuning offset or uncontrolled changes in the laser intensity. In numerical simulations and preliminary NSD-PV data, we observed that without the 1$^{\rm st}$ and the 2$^{\rm nd}$ parity state projection lasers, $\rm{L}_{P1}$ and $\rm{L}_{P2}$, sensitivity to systematic contributions resulting from the combination of $\mathcal{E}_{nr}$ and $\mathcal{B}$-field gradients increase dramatically, due to the retention of information about such fields from outside the region between these laser beams. Consequently, having high depletion efficiency for both lasers (i.e. a low OPR value) is of critical importance. We explored the effects of these imperfections by deliberately exaggerating them and then making NSD-PV measurements as outlined in Sec.~\ref{SysOverallValue}. Additionally, we investigated systematic effects resulting from the combination of unipolar $\mathcal{E}_{nr}$ pulses centered near the parity state projection laser application locations (i.e. at $z = z_{10}/z_{22}$ for $L_{\rm P1}$/$L_{\rm P2}$) and offsets in the corresponding laser detuning. 

\begin{figure}
\includegraphics[width=83mm]{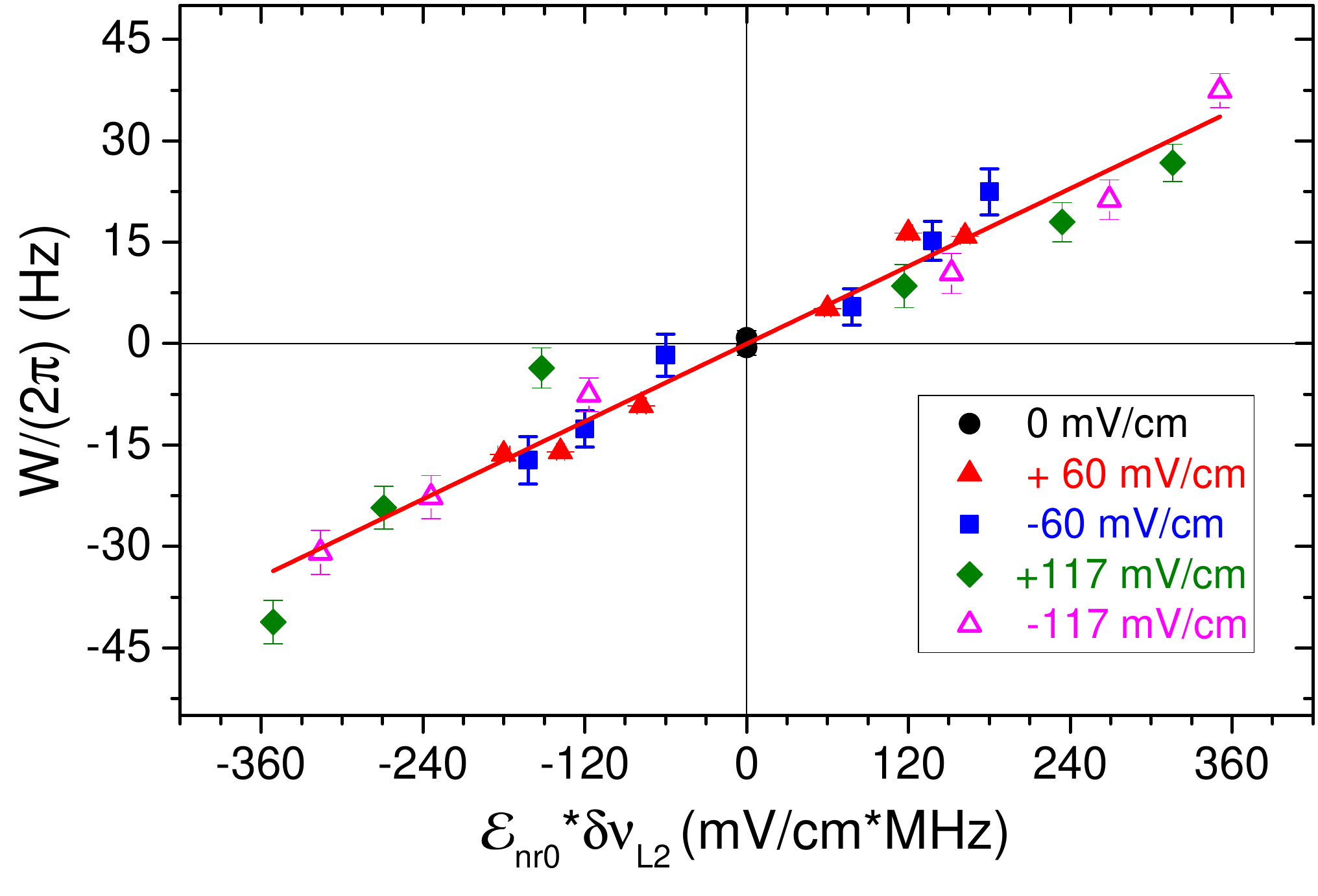}
\caption{(color online) (a) $W$ vs. $\mathcal{E}_{nr0} * \delta\nu_{L2}$ from measurements made with a deliberate offset in $L_{\rm P2}$ laser detuning,  $\delta\nu_{L2}$, and an intentional unipolar $\mathcal{E}_{nr}$ centered at Gap 22. The legend shows the value of $\mathcal{E}_{nr0}$ for each data point. The red line is a fit to the relation $W/(2\pi) = C_W * (\mathcal{E}_{nr0} * \delta\nu_{L2})$, yielding ${C_W=0.095\pm 0.004~\text{Hz/(MHz*mV/cm)}}$.}
\label{fig:Wa1_vs_EnrDetuning}
\end{figure}

We did not observe any systematic contribution to $W$ related to $\mathcal{E}_{nr}$ pulses applied near the $L_{\rm P1}$ beam. Hence, imperfections related to this laser beam are treated separately in the final systematic error budget. On the other hand, the combination of a unipolar $\mathcal{E}_{nr}$-field near the $L_{\rm P2}$ laser beam, $\mathcal{E}_{nr}^{\rm LP2}(z) = \mathcal{E}^u(z; z_{22})$, and a non-zero detuning offset in this laser, $\delta\nu_{L2}$, was observed to result in a non-zero shift in $W$. The resulting systematic offset in $W$ was found to be  proportional to the product of $\delta\nu_{L2}$ and the amplitude, $\mathcal{E}_{nr0}$, of the local non-reversing field near $z = z_{22}$ (see Fig.~\ref{fig:Wa1_vs_EnrDetuning}(a)). In additional data not shown here, we also observed that the systematic offset associated with this effect, $\Delta W_{\rm sys}^{\rm L2\mathcal{E}nr}$, is proportional to the ${\rm L}_{\rm P2}$ laser power.  Hence, the effect has the properties that would be expected for a DC Stark-induced AC Stark shift. However, to date we have not found a satisfactory model for an underlying physical mechanism that captures all observed features of this effect.  

This combination of imperfections also was found to result in a non-zero value for the parameter, $a_1$, used to describe a term in the asymmetry $\mathcal{A}(\Delta)$ that is linear in $\Delta$ (see Eqn.~\ref{Equ:DataFinalAsymFit}). We discuss the implications of this observation in Appendix~\ref{2ndDepA1}.

In order to determine the systematic error in $W$ resulting from this effect, we must determine the typical ambient laser detuning offset, $\Delta\delta\nu_{\rm L2}$, during NSD-PV runs. This is done as follows. First, by deliberately detuning the laser in known steps, we mapped out the dependence of the OPR on $\delta\nu_{\rm  L2}$ and fit this dependence to a simple functional form for OPR($\delta\nu_{\rm L2}$). Then, using the continuously recorded values of OPR during NSD-PV data runs, we inverted this relationship to determine the range of values of $\delta\nu_{\rm L2}$ during the run. The average OPR value for laser $L_{\rm P2}$ in a typical NSD-PV run is $8\%$, corresponding to a laser detuning offset of 
$\Delta \delta\nu_{\rm L2}=-0.3\pm1.3$ MHz.

We determine the systematic error in $W$ using the relationship established in Fig.\ref{fig:Wa1_vs_EnrDetuning}: 
\begin{equation}
W/(2\pi) = C_W * (\mathcal{E}_{nr0} * \delta\nu_{L2}), \label{eqn:Wa1_vs_EnrDelta3}
\end{equation}
where we found $C_W=0.095\pm 0.004$ Hz/(MHz*mV/cm). Here, $\mathcal{E}_{nr0}$ is the residual non-reversing $\mathcal{E}$-field near $L_{\rm P2}$ laser beam (i.e. $\mathcal{E}_{nr}(z= z_{22})$), whose typical mean and uncertainty under normal operation conditions are $\mathcal{E}_{nr0}^{a}(z_{22})=1.3\pm4.7$ mV/cm. Together with the typical ambient value for $\delta\nu_{L2}$ given above, the systematic error shift in $W$ and its associated uncertainty due to this combination of imperfections is $\Delta
W_{\rm sys}^{\rm{L}2\mathcal{E}nr}/(2\pi)=0.04\pm 0.21 ~\text{Hz}$.

\subsection{Total Systematic Shift and Uncertainty}
\label{SysFinal}

\begin{table*}
	\centering
\begin{tabular}{ p{5cm}  p{5.6cm}  p{3cm}  p{3cm}  }
		\hline
		\hline
\\[-0.6em]
Parameter & Ambient Value &Shift  \newline$(\Delta W)_{sys}/(2\pi)$ & Uncertainty $\delta W_{\text{sys}}/(2\pi)$ (Hz) \\
\hline	
\\[-0.6em]
	Bipolar $\mathcal{E}_{nr}$ Pulses & $\mathcal{E}_{nr0}^{a}=5.1~\text{mV/cm}$& &0.12\\
\\[-0.6em]	
Unipolar $\mathcal{E}_{nr}$ Pulses & $\mathcal{E}_{nr0}^{a}=5.1~\text{mV/cm}$ & &0.16 \\
\\[-0.6em]
	$\mathcal{B}$-Field Inhomogeneities & $\Delta^a/(2\pi)=0.21~\text{kHz}$ &  &0.24 \\
\\[-0.6em]
Linear $\mathcal{B}$-Field Gradient and & ${\partial{\Delta^{a}/(2\pi)}/\partial z= 0.03\pm0.02~\text{kHz/cm}}$, & -0.01 & 0.02\\
Unipolar $\mathcal{E}_{nr}$ at $z = z_{16}$&  ${\mathcal{E}_{nr0}^{a}(z_{16}) = - 1.5 \pm 3.5~\text{mV/cm}}$ &  & \\
\\[-0.6em]
	Detuning offset in $\rm L_{P2}$  and & $\Delta\delta\nu_{L2}=-0.3\pm1.3~\text{MHz}$,&-0.04  &0.21  \\
	$\mathcal{E}_{nr}$ at and near $z = z_{22}$ &  ${\mathcal{E}_{nr0}^{a}(z_{22})=1.3\pm4.7~~\text{mV/cm}}$ &  & \\
\hline
\\[-0.6em]
	Total Systematic & & -0.05 &0.38   \\
	\hline
	\hline 	 
\end{tabular}
\caption[Systematic shifts and uncertainties in $W$ and the total uncertainty.]{Systematic shifts and uncertainties in $W$ and the total uncertainty. All systematic errors are added in quadrature to obtain the total uncertainty in the systematic error in $W$.  }		
\label{Table:SysErr}
\end{table*}

We compute the systematic uncertainty in $W$ from the measurements with deliberately exaggerated parameters in the following way. As noted previously, we determine the amplification factor, $A_d=P^{\rm d}/P^{\rm a}$ for each type of exaggerated imperfection. For stray $\mathcal{E}$-fields, the RMS value of the typical $\mathcal{A}_{nr}(z)$ signal is defined to be the value for $P^a$; for $\mathcal{B}$-field inhomogeneities, the standard deviation of the typical measured $\mathcal{B}$-field under normal operating conditions is used. For the effect associated with a laser detuning offset, we determined the typical ambient laser detuning of $L_{\rm P2}$ laser as described above (Sec.~\ref{SysLaserDetuning}).

For each type of parameter, we calculate the weighted average, $W^d_{\rm avg}$, of $W^{+d}/A_d$ and $W^{-d}/[-A_d]$, and its uncertainty $\delta W^d_{\rm avg}$. Then, we define $(\delta W_{\text{sys}})^{\text{max}}$ --- the maximum value of the systematic error in $W$ associated with this parameter (same pulse shape applied at the same center location, etc.) --- as ${{\left(\delta W^d_{\text{sys}}\right)}^{\text{max}}}= |W_{\rm avg}^d+\text{sgn}(W_{\rm avg}^d)*\delta W_{\rm avg}^d|$. The total maximum systematic contribution for a given type of imperfection (e.g. $\mathcal{E}_{nr}$ fields, $\mathcal{B}$-field gradients, etc.) is computed by addition in quadrature of the values ${{\left(\delta W^d_{\text{sys}}\right)}^{\text{max}}}$ for each of the individual parameters of this type that we varied.

Only two combinations of imperfections were found to give definite shifts in $W$: (1) linear $\mathcal{B}$-field gradient and unipolar $\mathcal{E}_{nr}$ at $z = z_{16}$, and (2) non-zero detuning offset of $L_{\rm P2}$ laser plus $\mathcal{E}_{nr}$ near $z = z_{22}$. The systematic offset and its associated uncertainty are calculated as detailed in Sec.~\ref{GSLsystematics} for the first combination of parameters, and in Sec.~\ref{SysLaserDetuning} for the second combination.

Table~\ref{Table:SysErr} presents our total systematic error budget. Following the criteria outlined in Sec.~\ref{SysOverallValue}, we included contributions from 5 parameters in the calculation of the total systematic uncertainty. We find a total systematic shift, $\Delta W_{sys}/(2\pi) = -0.05$ Hz, and its associated uncertainty, $\delta W_{\text{sys}}/(2\pi)=0.38 $ Hz. We emphasize that our strategy of determining the total systematic uncertainty has been conservative, i.e., we included contributions from several parameters despite not observing direct manifestations of related systematic errors.

\subsection{Final Results}
\label{Conclusion}

\begin{table*}
	\begin{tabular}{ p{3cm}  p{2.5cm}  p{1.4cm}  p{3.cm}  p{4.1cm}  }
		\hline\hline
		\\[-0.9em]	
		Crossing  & $d$ (Hz/(V/cm)) & $\tilde{C}$  & $W/(2\pi)$ (Hz) & $W_{\rm mol}/2\pi =\kappa^{'}W_P/(2\pi)$ (Hz) \\ \hline
		\\[-0.8em]
		A	& 3360 & -0.41$i$    & $ 0.29\pm0.53\pm0.41$   & $ -0.71\pm1.29\pm1.00$ \\ 
		\\[-0.9em]
		F	& 3530 & +0.39$i$  & $0.00\pm0.55\pm0.41$	  & $0.00\pm1.41\pm1.05$  \\ 
		\\[-0.9em]
		Weighted Average	& -	& -   & -  & $-0.39\pm0.95\pm1.02$  \\ 	
		\hline \hline
	\end{tabular}
	\caption{Final results for all NSD-PV data with $^{138}\mathrm{BaF}$. The first uncertainty is statistical and the second systematic.}
	\label{table:WpKvalues}
\end{table*}

Table~\ref{table:WpKvalues} summarizes the weighted average results for each crossing. The experimentally measured values of the NSD-PV weak matrix element, $W$, in $^{138}$BaF at crossings A/F, (see Fig.~\ref{fig:Crossings138}) are
\begin{align}
W^{(A)}/(2\pi)=0.29\pm0.53\pm0.41~\text{Hz},\\
W^{(F)}/(2\pi)=0.00\pm0.55\pm 0.41~\text{Hz},
\end{align}
where the first uncertainty is statistical and the second systematic. Recall that the NSD-PV matrix element, $W$, connecting the nearly-degenerate levels is given by
$iW = \kappa' W_P \tilde{C}$,
where $W_P$ and $\kappa'$ have the same value at every crossing for a molecule with a given nucleus, whereas $\tilde{C}$ depends on the quantum numbers of the molecular states at the crossing. Hence, the matrix element $W$ has a different value at each crossing, but the quantity $W_{\text{mol}}\equiv\kappa' W_P$ should be the same at all crossings. Combining the NSD-PV measurements at the A and F crossings of $^{138}$BaF, we find $
W_{\text{mol}}/(2\pi)=-0.39\pm0.95\pm1.02~\text{Hz}$.
Here, since the systematic errors are evaluated for the entire set of measurements rather than for each crossing individually, we take the simple average of the individual systematic uncertainties (weighted by the statistical uncertainty for each crossing) as the final value for the total systematic uncertainty. However, because the statistical errors are not correlated, we average them in (weighted) quadrature in the usual way. Combining the statistical and the systematic errors in quadrature, we obtain our final result:
\begin{equation}
W_{\text{mol}}/(2\pi)\equiv\kappa' W_P/(2\pi)=-0.39\pm1.40~\text{Hz}.
\end{equation}

\subsection{Interpretation of the Result}
\label{Interpret}
In $^{138}$BaF, the calculated overlap of the valence electron with the $^{19}$F nucleus leads to $W_{P}(\rm F)\sim0.05$ Hz~\cite{Private}. Then, our NSD-PV measurement with $^{138}$Ba$^{19}$F can be interpreted as a measurement of the parameter $\kappa^\prime$ that quantifies the strength of the NSD-PV interactions of the $^{19}$F nucleus. We find
\begin{equation}
\kappa^{\prime}(^{19}\rm F)=-8\pm28.
\end{equation}  
This may be compared to a simple nuclear shell-model prediction for $\kappa^\prime(^{19}F)$, which we denote by $\kappa^\prime_{\rm thy}(^{19}\rm F)$. This is obtained by assuming that the $^{19}$F nucleus is described by a single valence proton in a $s_{1/2}$ orbital~\cite{Kopfermann}. This gives  $\kappa^\prime_{\rm thy}(^{19}\rm F) \approx -0.08$~\cite{Flambaum, Flambaum1984, DeMille}. The consistency of our measurement with this near-zero predicted value demonstrates the absence of systematic errors outside our range of uncertainty.

However, determining $\kappa^{\prime}(^{19}\rm F)$ was not the primary goal of this study. It is more useful to compare our demonstrated sensitivity to that of previous atomic PV experiments, and those projected for future molecular NSD-PV measurements. With $\sim\!80-90$ hours of data at each crossing, our statistical uncertainty for the NSD-PV matrix element for each crossing, $W^{(A)}$ or $W^{(F)}$, is $\delta W< 0.6$ Hz. The previous most sensitive atomic PV experiment, using Dy, had a statistical uncertainty of $2.9$ Hz with $\sim\!30$ hours of data~\cite{Nguyen}. In the next generation of our experiment, we plan to measure NSD-PV in $^{137}$BaF. The calculated value of $W_P$ in $^{137}$BaF is $W_{P}(^{137}\rm Ba) = 160 \pm 15$ Hz~\cite{Borschevsky, DeMille, Nayak,Labzowsky95,Isaev,Titov}. The crude expectation for the NSD-PV nuclear factor $\kappa^\prime$ is ${\kappa^\prime{(^{137}\rm Ba)}\approx\kappa^\prime_a+\kappa^\prime_2\approx0.07}$~\cite{DeMille}. Assuming the same statistical and systematic uncertainties as obtained in this work, the projected uncertainty, $\delta\kappa^\prime$, in $\kappa'{(^{137}\rm Ba)}$ would be $\delta\kappa'{(^{137}\rm Ba)}=\pm0.009$. This is hence expected to be sufficient for a measurement of $\kappa^{\prime}(^{137}\rm Ba)$ at the level of $\sim\! 10\%$. Our projected total uncertainty for measuring NSD-PV in $^{137}\rm BaF$ would represent a factor of ${\sim\!7}$ improvement on the sensitivity of the current best atomic measurement of NSD-PV: this experiment, using Cs atoms to measure the effect associated with the nucleus $^{133}$Cs, yielded $\kappa^{\prime}(^{133}\text{Cs})\approx0.39\pm0.06$~\cite{Wieman}. Based on our results here, the next generation of our experiment should not be limited by systematic effects, granted that the actual value of $\kappa^\prime$ is not much smaller than the predicted value.

A large number of molecular species with the required properties (level structure, known spectroscopic constants etc.) for measuring NSD-PV exist. Measuring the NSD-PV effect in several different species would make it possible to disentangle the different contributing effects~\cite{Piketty}. The overall size of the NSD-PV effect scales roughly as $(\kappa'_2 + \kappa'_a)Z^2$, where $Z$ is the atomic number. Hence, NSD-PV effects grow very rapidly with $Z$. The contribution from $Z^0$ exchange, $\kappa_2^\prime$, is independent of atomic mass number $A$, while the contribution from the nuclear anapole moment, $\kappa'_a$ scales roughly as $A^{2/3}$ ~\cite{Murray}. Hence, we can distinguish the effect due to the nuclear anapole moment from that due to $Z^0$ exchange with measurements over a range of nuclear masses. In heavier nuclei such as $^{137}$Ba, $^{69}$Ga, and $^{173}$Yb, the anapole moment term dominates, whereas in light nuclei like $^{9}$Be, $^{11}$B, or $^{47}$Ti, $Z^0$ exchange is expected to be primary~\cite{DeMille}.

Our technique is sufficiently general and already sensitive enough to enable measurements across a broad range of diatomic molecules. With planned improvements such as a much more intense and slower molecular beam source~\cite{BufferGas,Barry11,Hutzler11,Zhou15}, the sensitivity should even be sufficient to measure NSD-PV effects in light nuclei, where nuclear structure calculations are increasingly accurate. This gives the promise of using measurements of this type to provide a long-sought determination of purely hadronic PV interaction strengths ~\cite{Haxton01}. Future measurements with our technique also may be useful for constraining the strength of PV interactions mediated by lighter analogues of the $Z^0$ boson~\cite{Dzuba2017}.

\begin{acknowledgments}
This work was supported by the National Science Foundation under grant no. PHY-1404162.  We thank M.G.~Kozlov, D.~Murphree, D.A.~Rahmlow, E.~Kirilov, Y.V.~Gurevich, and R.~Paolino for contributions to earlier stages of this work.
\end{acknowledgments}

\appendix

\appendix
\section{Measuring the Non-Reversing $\mathcal{E}$-Field}
\label{Effects on Enr} 

\subsection{Extracted Versus Physical Fields}

By deliberately applying $\mathcal{E}_{nr}$-fields with known forms, it was observed that the functions $\mathcal{E}_{nr}(z)$ determined by simple application of Eqns.~\ref{Equ:EnrFinalDiff} and \ref{FourierSeriesE} had significant qualitative differences from the input fields.   To account for this, we draw a distinction between the actual, physical non-reversing field, $\mathcal{E}_{nr}(z)$, and the (inaccurate) non-reversing field we infer from the molecular data by simple application of  Eqns.~\ref{Equ:EnrFinalDiff} and \ref{FourierSeriesE}. We refer to this imperfect ``extracted'' field as $\mathcal{X}_{nr}(z; \{z_i\}; \mathcal{E}_{nr}(z))$, since we find that it depends on the set of values $\{z_i\}$ used as well as on the form of the underlying physical field $\mathcal{E}_{nr}(z)$. 

To understand the differences between the actual ($\mathcal{E}_{nr}$) and extracted ($\mathcal{X}_{nr}$) non-reversing fields, we made extensive studies of the response of the system to deliberately applied non-reversing field pulses, $\mathcal{E}_{nr}^{\rm del}(z;z_k)$, with the simple form of a unipolar field pulse centered at ring gap location $z_k$: $\mathcal{E}_{nr}^{\rm del}(z;z_k) = \mathcal{E}^{u}(z;z_k)$  (with peak amplitude $\mathcal{E}^{\rm del}_{nr0} = \mathcal{E}^u_0(\delta V = 0.7$ V)). For these studies,  we used a simple choice for the set of reversible field pulses: namely, we apply reversible field pulses $\mathcal{E}_{r}(z; z_i)$ to a set of 3 adjacent gap locations $\{z_{i-1},z_i,z_{i+1}\}$.  We denote such a triplet of gap locations for the reversible field pulses as  $\{\{z_i\}\}$.  The extracted fields under these conditions were found with experimental data, and also with simulated data generated by numerical solutions of the Schr\"{o}dinger equation (Eqn.~\ref{Equ:evenState1stOrder2}); both were subjected to the same extraction analysis.

\begin{figure}
\includegraphics[width=83mm]{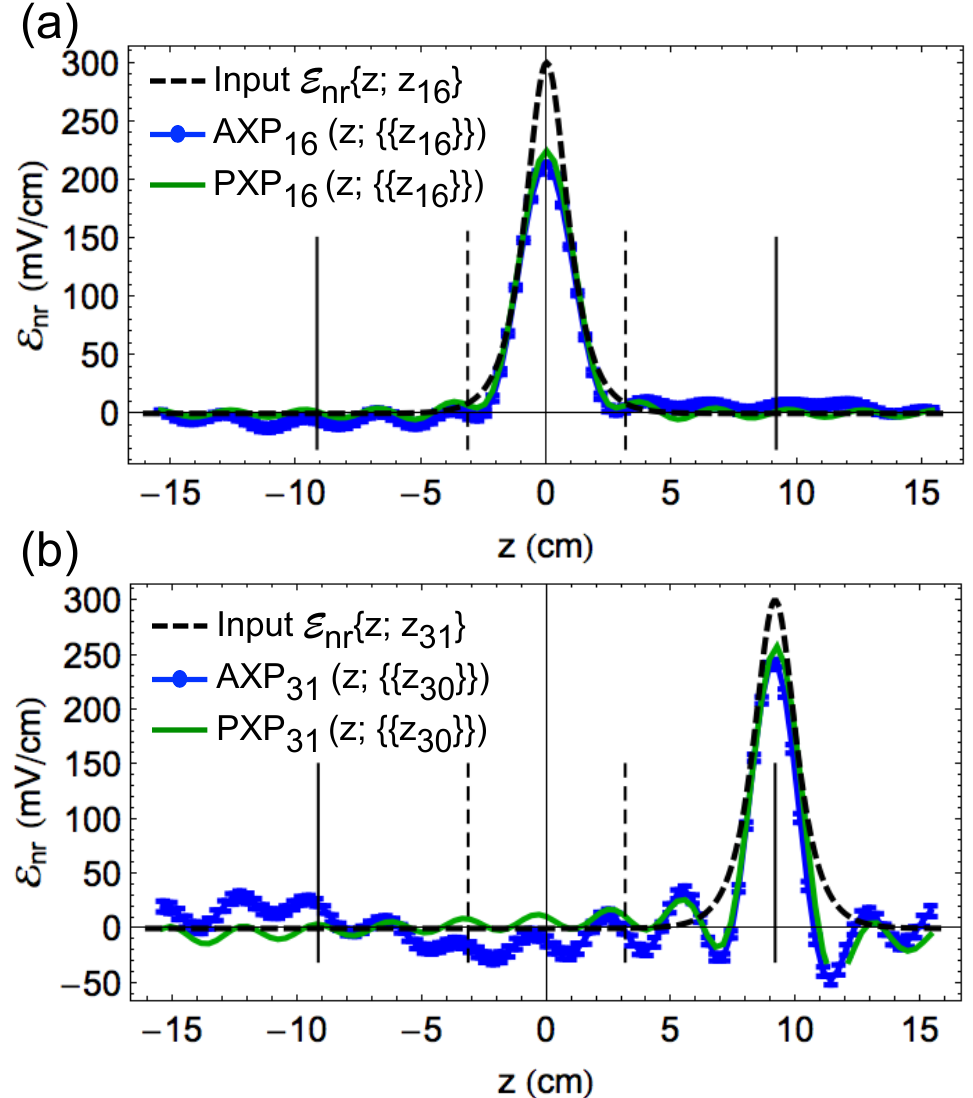}
\caption{(color online). Experimentally measured ``actual extracted field pulse'', $AXP_k(z;\{\{z_i\}\})$ (circles), and the corresponding ``perfect extracted field pulse'', $PXP_k(z;\{\{z_i\}\})$ (solid line), from numerical simulations. Dashed curves show the deliberately applied non-reversing field pulse, $\mathcal{E}_{nr}^{\rm del}(z; z_{k})$, used in both the numerical simulations and the experimental measurements. Both $AXP_k(z;\{\{z_i\}\})$ functions were measured with the condition $|k-i|\leq1$. (a) $AXP_{16}(z;\{\{z_{16}\}\})$ and $PXP_{16}(z;\{\{z_{16}\}\})$ compared for  $\mathcal{E}_{nr}^{\rm del}(z; z_{16})$. (b) $AXP_{31}(z;\{\{z_{30}\}\})$ and $PXP_{31}(z;\{\{z_{30}\}\})$ compared for $\mathcal{E}_{nr}^{\rm del}(z; z_{31})$.   Vertical dashed (solid) lines indicate $L_{\rm P1,2}$ laser beam locations (ends of the IR).} 
\label{fig:AXPandPXPcomparisons}
\end{figure}

In these studies, it became clear (in both the numerical simulations and the experimental data) that the finite range of detunings employed in our procedure led to significant spectral windowing artifacts.  The flat-top window implicit in our procedure (since we take data over a finite range of detunings and do not weight points differently) leads to significant broadening and sidelobes when transformed to the time/position domain (see Fig.~\ref{fig:AXPandPXPcomparisons}).  Attempts to use other windowing strategies did not lead to acceptable solutions. Instead, we used the following procedures to account for these effects.


As described earlier, we assume that $\mathcal{E}_{nr}(z)$ can be written as a superposition of unipolar field pulses $\mathcal{E}^u(z ;z_k)$, with associated weighting coefficients $c_k$.  Our field extraction procedure, described above, leads to a noticeably different result, the function we call $\mathcal{X}_{nr}(z; \{z_i\}; \mathcal{E}_{nr}(z))$.  However, all steps in the extraction procedure are linear in the fields of interest.  Hence, it should be the case that $\mathcal{X}_{nr}(z; \{z_i\})$ can be written as a superposition of the functions $\mathcal{X}(z; \{z_i\}; \mathcal{E}^{\rm del}_{nr}(z ;z_k))$ that result from applying the same extraction procedure to the unipolar field ``basis'' pulses.  In other words, these functions (the extracted versions of the input basis functions) can act as an effective basis set for any extracted function $\mathcal{X}_{nr}(z; \{z_i\})$.  Then, determining the coefficients $c_k$  such that 
\begin{equation}
\mathcal{X}_{nr}(z; \{z_i\}; \mathcal{E}_{nr}(z)) = \sum_{k=1}^{31} \mathcal{X}(z; \{z_i\}; \mathcal{E}^{\rm del}_{nr}(z ;z_k)),
\end{equation}
is sufficient for determining $\mathcal{E}_{nr}(z)$, since the same coefficients should appear in its expansion in terms of unipolar field pulses.

With this approach, the next task was to determine accurate representations of the ``extracted pulse basis functions'' $\mathcal{X}(z; \{z_i\}; \mathcal{E}^{\rm del}_{nr}(z ;z_k))$.  We noticed significant discrepancies between the measured and numerically simulated versions of these functions, except when the distance $|z_i-z_k|$ between the centers of the non-reversing field pulse ($z_k$) and the reversible field pulses (${z_i}$) was sufficiently small. (This was found to be due primarily to the finite longitudinal velocity spread of the molecular
beam. When averaged over the molecular ensemble, this leads to effective dephasing of the interfering amplitudes when $\mathcal{E}_{nr}$ and $\mathcal{E}_{r}$ are widely separated in time.) We found that, so long as $|k-i|\leq5$, the simulated and measured versions of the extracted fields agree out to $|z-z_k|\lesssim 2$ cm.   Then, to ensure that every position $z$ in the IR is probed with a reversible field pulse sufficiently near its location, we make final $\mathcal{E}_{nr}(z)$ measurements with 7 different sets $\{\{z_i\}\}$ of reversible pulse locations.  The list of these sets of gap triplets is given in Table~\ref{table:ErGapsForAnr}. This choice effectively divides the interaction region into 7 sections, such that reliable data on $\mathcal{E}_{nr}$ in each section is available from reversible field pulses applied in that section.  

\begin{table*}
	\centering
	\begin{tabular}{ p{3.2cm} | p{9cm} | p{4cm} }
		\hline\hline
		\\[-1em]
		Middle Gap Number of Applied $\mathcal{E}_{r}$ Set & Significance & Set of 3 Adjacent Locations  $\mathcal{E}_{r}$ Applied  \\ \hline 
		\\[-1em]  
		$i=2$ & First ring &$\{\{z_2\}\}\equiv \{z_{1}, z_2, z_{3}\}$
		\\ \hline
		\\[-1em]
		$i=5$ & Halfway between beginning of IR and $L_{P1}$ laser beam & $\{\{z_5\}\}\equiv \{z_{4}, z_5, z_{6}\}$\\ \hline
		\\[-1em]
		$i=10$ & Gap just before 1$^{\rm st}$ prism ring &$\{\{z_{10}\}\}\equiv \{z_{9}, z_{10}, z_{11}\}$\\ \hline
		\\[-1em]
		$i=16$ & Central electrode in the IR &$\{\{z_{16}\}\}\equiv \{z_{15}, z_{16}, z_{17}\}$\\ \hline
		\\[-1em]
		$i=21$ & Gap just before 2$^{\rm nd}$ prism ring & $\{\{z_{21}\}\}\equiv \{z_{20}, z_{21}, z_{22}\}$\\ \hline
		\\[-1em]
		$i=27$& Halfway between $L_{P2}$ laser beam and end of IR & $\{\{z_{27}\}\}\equiv \{z_{26}, z_{27}, z_{28}\}$\\ \hline
		\\[-1em]
		$i=30$& Last ring & $\{\{z_{30}\}\}\equiv \{z_{29}, z_{30}, z_{31}\}$
		\\ 	
		\hline \hline
	\end{tabular}
	\caption{The chosen set of 3 adjacent locations at which $\mathcal{E}_{r}(z; z_i)$ is applied during the sequence of measurements used to determine the extracted non-reversing field function, $\mathcal{X}_{nr}(z)$. }
	\label{table:ErGapsForAnr}
\end{table*}

We find it useful to define the ``actual extracted field pulses" $AXP_k(z;\{\{z_i\}\})$, i.e., the extracted field from experimental data with an applied unipolar ``basis set'' pulse centered at gap $k$ and determined by applying reversible pulses at the gap triplet $\{\{z_i\}\}$:  $AXP_k(z;\{\{z_i\}\}) = \mathcal{X}(z, \{\{z_i\}\};\mathcal{E}^{\rm del}_{nr}(z;z_k))$.
Ideally, we would measure $AXP_k(z;\{\{z_i\}\})$ at every gap $k=1,...31$ with a nearby reversible pulse gap triplet $\{\{ z_i \}\}$, and use these as the basis set for determining the coefficients $c_k$ that define $\mathcal{X}_{nr}(z)$ and $\mathcal{E}_{nr}(z)$.  However, these experimentally-determined functions are contaminated by noise and other imperfections. Moreover, generating this complete set of data would be even more time-consuming than taking a full set of NSD-PV data.

Instead, we used a limited set of experimental data of this type, in combination with numerically-simulated data, to generate a full set of effective basis functions for the extracted field.  This was done as follows.  We began with nine $AXP_k(z;\{\{z_i\}\})$ measurements, at $k=1,5,10,11,16,21,22,27,31$; for each, the reversible field pulse set $\{\{z_i\}\}$ was chosen from the set in Table~\ref{table:ErGapsForAnr} so that at least one member of the set satisfies the condition $|k-i| \leq 1$.  We refer to the set of values of $k$ for which $AXP_k(z)$ measurements were taken as $\{k\}_{\rm meas}$.  Next, we generated the analogous extracted field pulses, but now with numerically simulated data rather than actual experimental data, for all $\mathcal{E}^{\rm del}_{nr}(z;z_k)$ center locations ($k = 1-31$).  For each, the location of the reversible field pulse set  $\{\{z_i\}\}$ was chosen to be the closest to $z_k$, among the sets in Table~\ref{table:ErGapsForAnr}.  We refer to these functions extracted from simulated data as the ``perfect extracted field pulse" functions, $PXP_k(z;\{\{z_i\}\})$.  To (partially) account for any differences between these (noise-free) ``perfect'' extracted pulses and the actual experimental response of the system, we rescale the $PXP_k(z;\{\{z_i\}\})$ functions to make them match, as well as possible, the ``actual''  extracted pulses $AXP_k(z;\{\{z_i\}\})$.  These rescaled noise-free functions, referred to as  the "fitted extracted field pulse" functions $FXP_k\left(z; \{\{z_i\}\} \right)$, are defined as
\begin{equation}
FXP_k\left(z; \{\{z_i\}\} \right) = S_{k} \cdot  PXP_k\left(z; \{\{z_i\}\}\right),
\label{FittedFunc}
\end{equation}
where the  quantities $S_{k}$ are scaling coefficients.  We define their values as follows.
For values of $k$ in the set $\{k\}_{\rm meas}$, the scaling coefficients are chosen to minimize the mean squared deviation between the ``perfect'' and ``actual'' extracted pulses, i.e., such that 
\begin{equation}
\int \vert S_{k} \cdot  PXP_k\left(z; \{\{z_i\}\}\right) - AXP_k\left(z; \{\{z_i\}\} \right)\vert ^2 dz
\label{scalingCoeff}
\end{equation}
is minimized. 
For other values $k=k^\prime$, outside the set $\{k\}_{\rm meas}$, we set the scaling coefficient $S_{k^\prime}$ as follows.  We find the nearest value to $k^\prime$, within the set $\{k\}_{\rm meas}$, that shared the same set $\{\{z_i\}\}$ of reversible field pulses.  Denoting this nearest value as $k_{\rm near}$, we then set $S_{k^\prime} =S_{k_{\rm near}}$.

This final set of scaled functions $FXP_k\left(z; \{\{z_i\}\} \right)$, $k=1-31$, forms an effective non-orthonormal (but complete) basis set for the extracted field functions, which is both noise-free and the best match to experimentally-measured ``extracted'' response functions. Figure \ref{fig:AXPandPXPcomparisons} shows examples of the deliberately applied pulses $\mathcal{E}_{nr}^{\rm del}(z; z_{k})$ aw well as typical data for the corresponding extracted functions  $AXP_k(z;\{\{z_i\}\})$  and $PXP_{16}(z;\{\{z_{16}\}\})$.

\subsection{Assigned Non-Reversing Electric Field}
\label{Anr}

We ultimately determine the unknown ambient non-reversing field, $\mathcal{E}_{nr}(z)$, with the following procedure. We apply reversing unipolar field pulses at the gap triplets $\{\{z_i\}\}$ given in Table~\ref{table:ErGapsForAnr} and calculate the extracted non-reversing field $\mathcal{X}_{nr}(z;\{\{z_i\}\};\mathcal{E}_{nr}(z))$ with uncertainties $\delta\mathcal{X}_{nr}(z;\{\{z_i\}\};\mathcal{E}_{nr}(z))$.  We then write $\mathcal{X}_{nr}(z)$ as a superposition of the effective basis functions $FXP_k(z)$, with a set of associated weighting coefficients ${c_k}$.  The $c_k$ values and their uncertainties are determined by minimizing the deviation between $\mathcal{X}_{nr}(z)$ and the linear combination of $FXP_k(z)$ functions.  That is, we assign $c_k$ by minimizing the expression
\begin{equation}
\int \sum_i\frac{|\mathcal{X}_{nr}\left(z;\{\{z_{i}\}\}\right) -\sum_{k=1}^{31} c_{k} FXP_{k}\left(z; \{\{z_{i}\}\} \right)|^2}{\vert f\left( \vert z-z_i \vert \right)\cdot \delta\mathcal{X}_{nr}\left(z;\{\{z_{i}\}\}\right)  \vert^2 dz}. 
\label{eqn:AnrMinimize}
\end{equation}
Here the weighting function, $1/f \left( \vert z-z_i \vert \right)$,
is a simple analytic form chosen to account for the fact that when $|z-z_k|$ is too large, the discrepancy between ``actual'' and ``perfect'' field pulses is significant. We use $f \left( \vert z-z_i \vert \right) = \sqrt{1 + \left(\frac{\vert z-z_i \vert}{\Delta z}\right)^4}$, where $\Delta z= 0.5$ cm is chosen empirically to generate a good match between data and fits when deliberate non-reversing fields of the form $\mathcal{E}_{nr}^{\rm del}(z;z_k)$ are applied and then subjected to the extraction procedure.

Then, finally, we can determine the ``assigned non-reversing field", $\mathcal{A}_{nr}(z)$, which is our best estimate of the ambient non-reversing field determined from this complex procedure. This function is given by \begin{equation}
\mathcal{A}_{nr}(z) = \sum_{k=1}^{31} c_{k} \mathcal{E}^{\rm del}_{nr}(z; z_{k}),
\label{eQNAnr}
\end{equation} 
where the set of coefficients $\{ c_k \}$ is determined as described above.


\section{2$^{\rm nd}$ Parity State Projection Laser Imperfections and $a_1$ Fit Parameter}
\label{2ndDepA1}

\begin{figure}
\includegraphics[width=83mm]{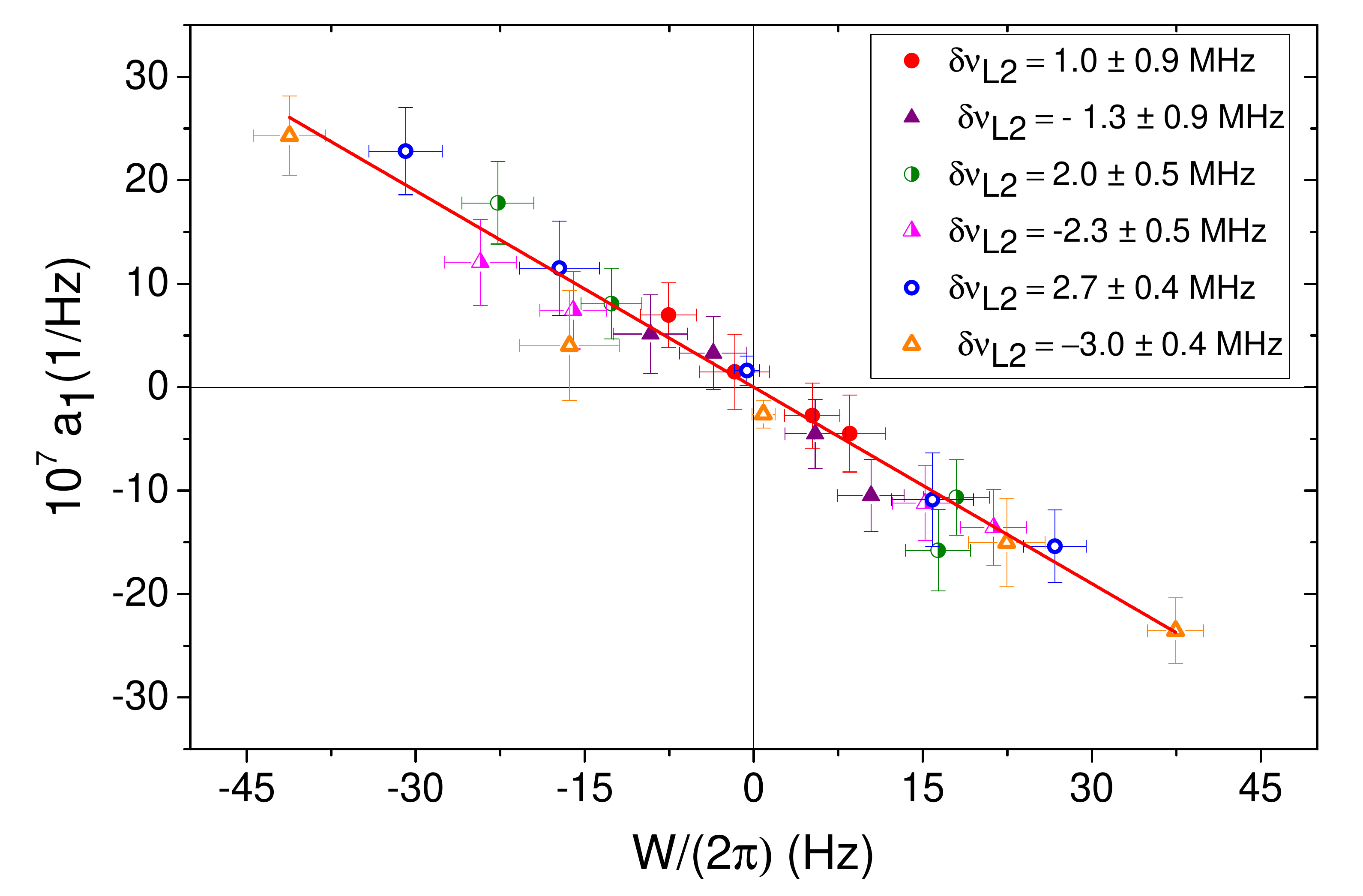}
\caption{(color online) Plot of $a_1$ vs. $W$ from measurements made with a deliberate offset in $L_{\rm P2}$ laser detuning,  $\delta\nu_{L2}$, and an intentionally amplified unipolar $\mathcal{E}_{nr}$ centered at $z = z_{22}$. The red line is a fit to the linear relationship $a_1=b_1 W$. This yields ${b_1=(-6.4\pm0.4)*10^{-8}~\text{1/Hz}^2}$ . }
\label{fig:a1_vs_EnrDetuning}
\end{figure}

We observed a clear linear relationship between $a_1$ and $W$ (Fig.~\ref{fig:a1_vs_EnrDetuning}) in measurements with the combination of a unipolar $\mathcal{E}_{nr}$-field near $L_{\rm P2}$ laser beam (i.e. near $z=z_{22}$), $\mathcal{E}_{nr}^{u}(z;z_{22})$, and a non-zero detuning offset in this laser, $\delta\nu_{L2}$. Thus, a non-zero $a_1$ fit result in a regular NSD-PV run (i.e. without any parameters deliberately exaggerated) could be treated as a preliminary indication of this type of systematic error as well. This correlation between $a_1$ and $W$ provides us with an additional and independent method of checking the systematic contribution to $W$. Using the observed relationship $a_1 = b_1 W$, we project the maximum systematic contribution to $W$, $W_{\rm sys}^{\rm max}$, is $W_{\rm sys}^{\rm max}=a_1^{\rm avg}/b_1$, where $a_1^{\rm avg}$ is the average $a_1$ fit result from NSD-PV runs given in Figure~\ref{fig:AllPVonlyA0BothCrossings} and $b_1$ is the slope of the line in Figure~\ref{fig:a1_vs_EnrDetuning}. The values of $W_{\rm sys}^{\rm max}$ for each crossing, deduced using this approach, are
\begin{align}
	(W_{\rm sys}^{\rm max})^{\rm A}/(2\pi)&=0.3\pm1.0~\text{Hz},\\
	(W_{\rm sys}^{\rm max})^{\rm F}/(2\pi)&=0.4\pm1.0~\text{Hz},
\end{align}
each consistent with the null result. The method described in the main text for evaluating this contribution to the systematic error yields smaller uncertainties. Hence, we treat this evaluation of $W_{\rm sys}^{\rm max}$ as a preliminary check for systematic errors related to $\delta\nu_{L2}$ and $\mathcal{E}_{nr}^{u}(z_{22})$, and use the calculations described in Sec.~\ref{SysFinal} to assign the total systematic uncertainty in $W$.

\bibliography{PRA_138BaF_22Dec17}

\end{document}